\documentclass[10pt,journal,letterpaper]{main}

\ifCLASSINFOpdf
   \usepackage[pdftex]{graphicx}
  \usepackage{mathrsfs}
  \usepackage{color}
  \usepackage{amsthm}
    \usepackage{soul}
    \soulregister\cite7 
    \soulregister\ref7
  \usepackage{amsmath}
  \usepackage[square, comma, sort&compress, numbers]{natbib}
  \usepackage{algorithm}
  \usepackage{algorithmicx}
  \usepackage{algpseudocode}

  \usepackage{enumerate}

  \usepackage[square,numbers,sort&compress]{natbib}

  \newcommand{\tabincell}[2]{\begin{tabular}{@{}#1@{}}#2\end{tabular}}

  \usepackage{array}
  \newcommand{\PreserveBackslash}[1]{\let\temp=\\#1\let\\=\temp}
  \newcolumntype{C}[1]{>{\PreserveBackslash\centering}p{#1}}
  \newcolumntype{R}[1]{>{\PreserveBackslash\raggedleft}p{#1}}
  \newcolumntype{L}[1]{>{\PreserveBackslash\raggedright}p{#1}}

  \usepackage{marvosym}
  \usepackage{CJK}
  \usepackage{amsfonts}
  \usepackage{subfigure}

  \usepackage{multirow}

  \def\mbG{{\mathbb G}}
  
  \def\mbI{{\mathbb I}}
  \def\mbT{{\mathbb T}}
  
  \def\mbZ{{\mathbb Z}}
  \def\mbC{{\mathbb C}}

  \newtheorem{myDef}{Definition}

\else
\fi

\usepackage{fancyhdr}

\begin{document}



\title{Secure Phrase Search for Intelligent Processing of Encrypted Data in Cloud-Based IoT}

\author{Meng Shen,~\IEEEmembership{Member,~IEEE,}
        Baoli Ma,
        Liehuang Zhu,~\IEEEmembership{Member,~IEEE,}\\
        Xiaojiang Du,~\IEEEmembership{Senior Member,~IEEE,}
        and Ke Xu,~\IEEEmembership{Senior Member,~IEEE}
\IEEEcompsocitemizethanks{
\IEEEcompsocthanksitem This work is partially supported by the National Key Research and Development Program of China under Grant 2018YFB0803405, the National Natural Science Foundation of China under Grants 61602039 and 61472212, the EU Marie Curie Actions CROWN under Grant FP7-PEOPLE-2013-IRSES-610524, and CCF-Tencent Open Fund WeBank Special Funding.
\IEEEcompsocthanksitem M. Shen, B. Ma and L. Zhu are with Beijing Engineering Research Center of High Volume Language Information Processing and Cloud Computing Applications, School of Computer Science, Beijing Institute of Technology, Beijing 100081, China. Email: \{shenmeng, baolimasmile, liehuangz\}@\}@bit.edu.cn. Prof. L. Zhu is the corresponding author.
\IEEEcompsocthanksitem X. Du is with the Department of Computer and Information Sciences, Temple University, Philadelphia PA19122, USA. Email: dxj@ieee.org.
\IEEEcompsocthanksitem K. Xu is with the Department of Computer Science and Technology, Tsinghua University, Beijing 100084, China. Email: xuke@mail.tsinghua.edu.cn.
\IEEEcompsocthanksitem Copyright\copyright 2012 IEEE. Personal use of this material is permitted. However, permission to use this material for any other purposes must be obtained from the IEEE by sending a request to pubs-permissions@ieee.org.
}
}

\maketitle


\begin{abstract}
Phrase search allows retrieval of documents containing an exact phrase, which plays an important role in many machine learning applications for cloud-based IoT,
such as intelligent medical data analytics.
In order to protect sensitive information from being leaked by service providers, documents (e.g., clinic records) are usually encrypted by data owners before being outsourced to the cloud.
This, however, makes the search operation an extremely challenging task.
Existing searchable encryption schemes for multi-keyword search operations fail to perform phrase search, as they are unable to determine the location relationship of multiple keywords in a queried phrase over encrypted data on the cloud server side.

In this paper,
we propose \texttt{P3}, an efficient privacy-preserving phrase search scheme for intelligent encrypted data processing in cloud-based IoT.
Our scheme exploits the homomorphic encryption and bilinear map to determine the location relationship of multiple queried keywords over encrypted data.
It also utilizes a probabilistic trapdoor generation algorithm to protect users' search patterns.
Thorough security analysis demonstrates the security guarantees achieved by \texttt{P3}.
We implement a prototype and conduct extensive experiments on real-world datasets.
The evaluation results show that compared with existing multi-keyword search schemes, \texttt{P3} can greatly improve the search accuracy with moderate overheads.
\end{abstract}



\section{Introduction}

\IEEEPARstart{P}{hrase} search, which allows users to search for sentences or documents containing a specific phrase that consists of a set of consecutive keywords \cite{2014_Phrase_query_optimization_on_inverted_indexes},
serves as an important building block in many machine learning applications for cloud-based IoT \cite{tifs-shen2017}.
For instance, it can be applied to intelligent clinical data analytics collected from medical IoT devices,
which retrieves medical records related to a certain disease (e.g., myocardial infarction) and feeds machine learning algorithms to obtain portent symptoms of the disease.
It can also be applied to the emerging entity-oriented search \cite{entity}, which identifies the records within which the exact description of an \emph{entity}
(e.g., person or event) occurs.
The resulting records can be utilized for situation assessment and intelligent decision making.
Another application scenario refers to the semantic search in knowledge graphs, which searches for entities with semantic similarity (e.g., titles, positions, and interests) and provides input signals to machine learning models for recommendation of products, news, and advertisements.

The combination of cloud computing and IoT enables powerful processing of data beyond individual IoT devices with limited  capabilities.
This, however, raises a great concern about the security and privacy of IoT data stored in the cloud,
as untrusted cloud service providers may get access to sensitive data or even result in data leakage accidents \cite{shen2018content,tifs-shen2018}.
In order to protect data privacy,
data owners can opt to encrypt their sensitive data before outsourcing the storage of the data to remote cloud servers.
For instance, a healthcare company may store their encrypted patients' records in the cloud,
and allow only the authorized users to perform phrase search over these records.
This naturally imposes a requirement on the cloud-based search engine to perform phrase search operations over encrypted data.


Many schemes \cite{Fuzzy-Keyword-Search-over-Encrypted-Data-in-Cloud-Computing,
first:single:keyword:ranking_search,
Public-Key-Encryption-with-Keyword-Search,
Ananthi2011Privacy,
Yang2013A,
Fu2016Enabling,
Li2015Enabling,
Xia2016A,
Liu2016Privacy,
Li2016Enabling,
Fu2017Toward,
multi:keyword:ranked:search, multi:keyword:ranked:search:poly-func,
Privacy-Preserving_Multi-Keyword_Supporting_Ranking,
multi-keyword:inverted:index,
Multi-Keyword-Fuzzy-Search-2014,
An:Efficient:Privacy:Preserving:Ranked:Keyword:Search:Method,
first:phrase:search,
seaond:phrase:search,
A:Low:Storage:Phase:Search:Scheme:Based:on:Bloom:Filters:for:Encrypted:Cloud:Services,
Curtmola:2006:SSE:1180405.1180417,
Dynamic_Searchable_Symmetric_Encryption}
have been proposed to enable efficient search operations over encrypted textual data, as summarized in Table \ref{tab:summary}.
Existing solutions to the single-keyword and multi-keyword search problems cannot be used to perform phrase search over encrypted documents,
because they are unable to determine the positional\footnote{We use the terminologies of \emph{positional information} and \emph{location information} interchangeably in this paper.} relationship of the keywords composing a phrase in the encrypted environment.
For instance, the conjunctive keyword search scheme \cite{multi:keyword:ranked:search} will return a document if it contains each keyword at least once, regardless of whether these keywords appear consecutively as a phrase.
Therefore, if we use this scheme for phrase search, we would end with inaccurate results (cf. Section \ref{sec:evaluation}).

There are a limited number of studies targeting the phrase search problem over encrypted data
\cite{first:phrase:search,
seaond:phrase:search,
A:Low:Storage:Phase:Search:Scheme:Based:on:Bloom:Filters:for:Encrypted:Cloud:Services}.
These solutions, however, generally involve notable limitations as shown in Table \ref{tab:summary}, e.g.,
by either requiring resource-consuming multiple rounds of client-server interactions, or relying on a trusted third-party (TTP) for search result refinement on the behalf of the client.

Since the client-side IoT devices usually have constrained computing and storage resources, we aim at developing a phrase search scheme that achieves all of the attributes listed in Table \ref{tab:summary}.
The main challenge is to enable cloud servers to make a judgement on whether the keywords occurring in an encrypted document are consecutive or not,
without leaking sensitive information.

In this paper, we propose \texttt{P3}, a new \underline{P}rivacy-\underline{P}reserving \underline{P}hrase search scheme over cloud-based encrypted data.
We take advantage of the inverted index structure to build a secure index that achieves greater flexibility and efficiency.
The inverted index is one of the most popular and efficient index structures for plaintext search.
Compared with the diverse self-designed index structures \cite{multi:keyword:ranked:search, multi:keyword:ranked:search:poly-func,
Privacy-Preserving_Multi-Keyword_Supporting_Ranking, Multi-Keyword-Fuzzy-Search-2014,
An:Efficient:Privacy:Preserving:Ranked:Keyword:Search:Method},
the inverted index structure can improve retrieval efficiency and scalability in practice.
To tackle the challenge of determining the positional relationship of queried keywords over encrypted data,
we resort to the homomorphic encryption and bilinear map,
which enables the client to obtain exact search results from a single interaction with the cloud server.
As the phrase search is a special case of multi-keyword search,
our solution can also perform conjunctive multi-keyword search efficiently.

The main contributions of this paper are as follows:
\begin{enumerate}
  \item We propose a secure single-interaction phrase search scheme that enables phrase search over encrypted data in cloud-based IoT, without relying on a trusted third-party.
  \item We employ the combination of homomorphic encryption and bilinear map to determine the pairwise positional relationship of queried keywords on the cloud server side. It can be used as a building block in other relevant application scenarios.
  \item We implement a prototype of \texttt{P3} and conduct extensive experimental evaluation using real-world datasets. Results demonstrate that  \texttt{P3} greatly improves the search
accuracy with moderate overheads.
\end{enumerate}

The rest of paper is organized as follows.
We summarize the related work in Section \ref{sec:related_work} and present the problem formulation in Section \ref{sec:PROBLEM FORMULATION}.
We describe the proposed scheme in Section \ref{sec:main scheme}
and provide the security analysis in Section \ref{sec:security}.
We evaluate \texttt{P3} through extensive experiments in Section \ref{sec:evaluation} and discuss the limitations in Section \ref{sec:discussion}.
Finally, we conclude this paper in Section \ref{sec:conclusion}.

\section{Related Work}\label{sec:related_work}
The privacy-preserving data processing problem has attracted great research attention during the last decade \cite{jsac-zhu, network-gao,du_info13,Li2018}.
The secure searchable encryption problem was first addressed by Song et al. \cite{first_keyword_search},
which was index-free and could merely support exact single keyword search.
In order to extend the functionality and efficiency of searchable encryption,
follow-ups have proposed various schemes that support single keyword search \cite{Curtmola:2006:SSE:1180405.1180417,
Fuzzy-Keyword-Search-over-Encrypted-Data-in-Cloud-Computing, first:single:keyword:ranking_search} and exact or fuzzy multi-keyword search \cite{multi:keyword:ranked:search, multi:keyword:ranked:search:poly-func, Multi-Keyword-Fuzzy-Search-2014,
multi-keyword:inverted:index, Privacy-Preserving_Multi-Keyword_Supporting_Ranking,
An:Efficient:Privacy:Preserving:Ranked:Keyword:Search:Method,
Fu2017Toward,
Li2015Enabling,
Xia2016A,
Li2016Enabling,
Liu2016Privacy},
by using either self-designed indexes or the typical inverted index structure.
Several attempts have been taken to extend the fuzzy multi-keyword search scheme to support phrase search,
either by treating a pre-defined phrase (e.g., network security) as a single keyword \cite{Privacy-Aware-BedTree-Based-Solution} or introducing a TTP server on the client side \cite{first:phrase:search}.

Tang et al. \cite{seaond:phrase:search} proposed a phrase search construction over encrypted cloud data, but failed to implement and evaluate their proposal in real-world application scenarios.
For each individual phrase recognition,
this construction needed two rounds of communications between the client and the server, and also required a large number of trapdoors generated by the client.
The authors in \cite{A:Low:Storage:Phase:Search:Scheme:Based:on:Bloom:Filters:for:Encrypted:Cloud:Services}
proposed a phrase search scheme with relatively low storage and computational overhead.
However, they failed to present a complete threat model, a security definition, or a reasonable security proof.
Therefore, it remains unclear about the privacy guarantees provided by the proposed method.

\begin{table}[!t]
\renewcommand{\arraystretch}{1}
\caption{Summary of prior solutions and P3} \label{tab:summary}
\centering
\footnotesize{
\begin{tabular}{|c|c|c|c|c|}
\hline
Solutions      & \tabincell{c}{Multiple \\ Keywords}
& \tabincell{c}{ Phrase \\ Search }
& \tabincell{c}{ Single \\ Interaction }
& \tabincell{c}{ T.T.P. \\ Free } \\ \hline
\tabincell{c}{
Single keyword \\
search \cite{Curtmola:2006:SSE:1180405.1180417,
Fuzzy-Keyword-Search-over-Encrypted-Data-in-Cloud-Computing,
first:single:keyword:ranking_search}
} & $\times$ & $\times$  & $\surd$ & $\surd$\\ \hline
\tabincell{c}{
Multi-keyword search \\
\cite{multi:keyword:ranked:search, multi:keyword:ranked:search:poly-func,
Privacy-Preserving_Multi-Keyword_Supporting_Ranking, multi-keyword:inverted:index}
\\ \cite{Multi-Keyword-Fuzzy-Search-2014,
An:Efficient:Privacy:Preserving:Ranked:Keyword:Search:Method,
Fu2017Toward,
Xia2016A,
Liu2016Privacy}
} & $\surd$ & $\times$ & $\surd$ & $\surd$\\ \hline
\tabincell{c}{
Phrase search \\ \cite{seaond:phrase:search,
A:Low:Storage:Phase:Search:Scheme:Based:on:Bloom:Filters:for:Encrypted:Cloud:Services}
}& $\surd$ & $\surd$       & $\times$    &   $\surd$             \\ \hline
\tabincell{c}{
Phrase search \\ \cite{first:phrase:search}
} &
$\surd$ & $\surd$ & $\surd$ & $\times$  \\ \hline
\texttt{P3}        & $\surd$            & $\surd$     & $\surd$    &  $\surd$    \\
\hline
\end{tabular}}
\end{table}

In contrast to the existing phrase search solutions,
the phrase search scheme proposed in this paper
is a single-interaction scheme without a TTP.
Therefore, it can achieve higher flexibility and lower communication overhead.

\section{Problem Formulation}\label{sec:PROBLEM FORMULATION}
In this section,
we formally define the secure phrase search problem in intelligent processing of encrypted data.
We denote several keywords whose locations in the documents are consecutive are a \emph{phrase}.
We denote a keyword collection of the documents and their corresponding document identifier and location information
as an \emph{index}, and an encrypted index as a \emph{secure index}.
We refer to a searched phrase as a \emph{query} and an encrypted query as a \emph{trapdoor}.

\subsection{System Model}
\begin{figure}[t]
\begin{minipage}{0.49\textwidth}
  \centering
  \includegraphics[height=3.5cm]{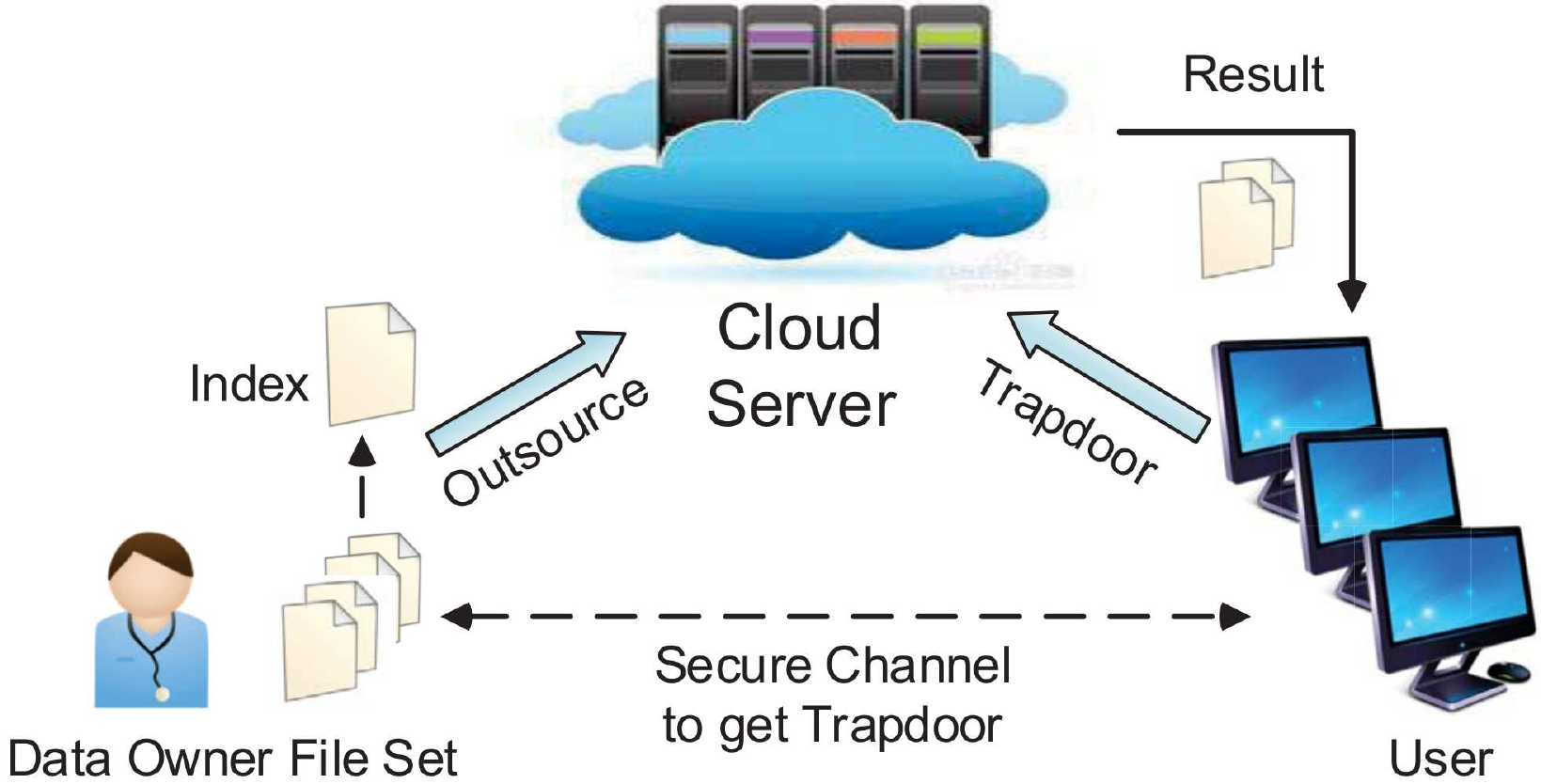}
\end{minipage}
\caption{System model of cloud-based phrase search over encrypted data}\label{fig:system_model}
\end{figure}

The privacy-preserving phrase search system over encrypted data involves three entities, namely
a \emph{IoT data owner}, a \emph{cloud server}, and one or multiple \emph{users},
as illustrated in Fig. \ref{fig:system_model}.
The data owner generates a secure searchable index for the document set and outsources the secure index along with the encrypted document set to the cloud server.
When an authorized user, say Alice, performs a phrase search over the encrypted documents,
she first acquires the corresponding trapdoor from the data owner through the search control mechanism (e.g., broadcast encryption \cite{Curtmola:2006:SSE:1180405.1180417}), and then submits the trapdoor to the cloud server.
Upon receiving Alice's trapdoor, the cloud server executes the predesigned search algorithms and replies to the user with the corresponding set of encrypted documents as the search results.
Finally, the user decrypts the received documents with the help of the data owner.

We assume that both the user and the data owner have limited computation and storage capacities on a practical basis.
Existing key management mechanisms
\cite{du_twc09,du_survey,Du_adhoc_07}
can be employed to manage the encryption capabilities of authorized users.

The above scheme is formally defined as follows:

\begin{myDef}
(Privacy-Preserving Phrase Search Scheme).
\emph{
A privacy-preserving phrase search scheme consists of the following polynomial time algorithms:
\begin{itemize}
  \item \textbf{KeyGen$(\tau, d)$:}
  Let $\tau$ and $d$ be security parameters as inputs of KeyGen$(\cdot)$, and a master key $Mk$ be an output.
  \item \textbf{IndexGen$(Mk, \Gamma)$:}
  It executes on the data owner side and takes the master key $Mk$ and the document collection $\Gamma$ as inputs and the secure index $\widehat{\mbI}$ as an output.
  \item \textbf{TrapdoorGen$(Mk,Q)$:}
  Given the master key $Mk$ and a query $Q$ from a user, it outputs the secure trapdoor ${\mbT}_{Q}$.
  This process is also performed on the data owner side.
  \item \textbf{Query$(\widehat{\mbI}, {\mbT}_{Q})$:}
  Given the secure index $\widehat{\mbI}$ and the trapdoor ${\mbT}_{Q}$, it performs search operations on the cloud server side and returns query results.
\end{itemize}
}
\end{myDef}

\subsection{Security Model}
Similar to the existing searchable encryption solutions
\cite{multi-keyword:inverted:index,
Multi-Keyword-Fuzzy-Search-2014},
we consider the cloud server
as an \emph{honest-but-curious} adversary.
That is, the cloud server would honestly follow the predesigned phrase search protocols and correctly provide the corresponding services to users,
but, it may be curious about the contents of the documents and attempt to learn additional information by analyzing the trapdoor and indexes.
For instance, it would infer the keywords in the index and trapdoors, as well as their locations in the documents.

Motivated by the existing literature \cite{multi:keyword:ranked:search, multi:keyword:ranked:search:poly-func, Multi-Keyword-Fuzzy-Search-2014},
we consider the following two threat models with different attack capabilities, depending on the sensitive information that can be obtained by the cloud server:
\begin{itemize}
  \item \textbf{Known Ciphertext Model.}
  The cloud server can only access the encrypted document set and the corresponding secure index that are outsourced by the data owner, and the trapdoors submitted by users.
  The cloud server is also capable of recording the search history, such as the search results in terms of encrypted documents.

  \item \textbf{Known Background Model.}
  In this stronger model, the cloud server is assumed to be aware of more facts than what can be known in the
  \emph{known ciphertext model}.
  In particular, the cloud server can learn the statistical information, such as keyword frequency in the document set.
  Furthermore, given such statistical information, the cloud server may infer the keywords in a queried phrase.
\end{itemize}

Our scheme aims at protecting privacy associated with the phrase search operation, which consists of three types of privacy, namely the document set privacy,
the index privacy, and the trapdoor privacy.
The document set privacy can be easily achieved by encrypting the documents using a block cipher,
such as AES, before outsourcing them to the cloud server.
Therefore, in this paper we focus on the latter two aspects,
which are described as follows:
\begin{itemize}
  \item \textbf{Index privacy}.
  Since the secure index can be regarded as a representation of the encrypted documents,
  any further information (e.g., keywords) should not be deduced from the index by the cloud server,
  except for the relationship between a trapdoor and its corresponding search results.
  In general, index privacy refers to the information of keywords, document identifiers, and keyword locations.
  Here, the keyword location privacy is guaranteed once the location information of all keywords is protected.
  We assume that the relationship between the keyword locations can be revealed to the cloud server, which does not go against the keyword location privacy.

  \item \textbf{Trapdoor unlinkability}.
  The trapdoors are used by the cloud server to perform matches with the secure index.
  Intuitively, the trapdoors should not reveal any valuable information (e.g., search frequency).
  The unlinkability means that the cloud server is unable to associate a trapdoor with the corresponding search phrase, i.e., the trapdoors generated for the same plaintext phrase should be different in multiple queries (e.g., queries submitted by multiple users or at different time periods).
\end{itemize}

%
%

\subsection{Definition and Notation} \label{subsec:notations}
Now, we introduce the main notations and the rest of the notations are summarized in Table \ref{tab:notations}.
\begin{itemize}
  \item $\Gamma$: a finite set of documents stored in plaintext, denoted as $\Gamma=(f_{1}, f_{2}, \dots,f_{m})$, where $f_{i}$ is the $i$-th document.

  \item $W$: a finite set of keywords extracted from the document set $\Gamma$,
  denoted as $W=(w_{1}, w_{2}, \dots, w_{\mu})$, where $w_{i}$ is the $i$-th keyword in $W$.

  \item $\mbI$: an inverted index of the document set $\Gamma$, denoted as $\mbI=({\mbI}_{w_{1}}, {\mbI}_{w_{2}}, \dots, {\mbI}_{w_{\mu}})$,
  where ${\mbI}_{w_{i}}$ is the inverted list corresponding to $w_{i}$.
 For each inverted list, we have ${\mbI}_{w_{i}}=(w_{i}, \Omega_{i1}, \Omega_{i2}, \dots, \Omega_{ik})$,
 where $\Omega_{ij}$ represents the $j$-th entity in ${\mbI}_{w_{i}}$. Let $\Omega_{ij} = (f_{ij}, \Lambda_{ij})$ be a tuple of the document identifier $f_{ij} \in \Gamma$ (${1} \le j \le {k}$) and the location identifier $\Lambda_{ij}$.
 $\Lambda_{ij}$ is a list of keyword locations in
 $f_{ij}$, which is denoted by $\Lambda_{ij} = \langle l_{j1}, l_{j2}, \dots, l_{jt} \rangle $.
 Here $l_{jr}$ ($1\le{r}\le{t}$) is the location where the keyword $w_{i}$ appears in the document $f_{ij}$.
\end{itemize}

\begin{table}[t]
\renewcommand{\arraystretch}{1.3}
\caption{Notations for Phrase Search Scheme}\label{tab:notations} \centering
\footnotesize{
\begin{tabular}{|c|l|}
\hline
Notation                  & Definition \\ \hline
$\widehat{\mbI}$          & The encrypted form of $\mbI$\\
$Q$                       & A query consisting of multiple keywords in $W$\\
$|Q|$                     & Number of keywords in the query $Q$\\
${\mbT}_{Q}$              & Trapdoor of the query $Q$\\
$|\mbI|$                  & Number of items in the inverted index\\
$|{\mbI}_{w_{i}}|$        & Number of documents containing the keyword $w_{i}$ \\
$\eta$                    & Maximum number of entries in a secure inverted list\\
\hline
\end{tabular}}
\end{table}

\subsection{Preliminaries}\label{subsec:Preliminaries}

\emph{\textbf{Bilinear map}} is a function combining elements of two groups (e.g., ${\mbG}_{1}$ and ${\mbG}_{2}$) to yield an element of a third group (e.g., ${\mbG}_{T}$).
We now briefly review it. For simplicity, we consider a special case where ${\mbG}_{1}={\mbG}_{2}={\mbG}$.

Let ${\mbG}$ and ${\mbG}_{T}$ be two (multiplicative) cyclic groups of a finite order $n$, and $g$ be a generator of ${\mbG}$. A bilinear map $e$ is a function in Eq. \eqref{eq:bilinear_map},
\begin{equation}\label{eq:bilinear_map}
e:{\mbG}\times{\mbG}\to{{\mbG}_{T}},
\end{equation}
with a useful property:
for all $u,v \in{\mbG}$ and $a,b \in{\mbZ}$,
we have $e(u^{a},v^{b})=e(u,v)^{ab}$,
yet $e(g,g)$ is a generator of ${\mbG}_{T}$.


\emph{\textbf{Homomorphic encryption}} is a cryptography primitive that allows us to perform operations over encrypted data
without knowing the secret key or decrypting the data.
Boneh et al. \cite{boneh2005evaluating} proposed a homomorphic encryption scheme based on
finite groups of composite order that supported a bilinear map,
which can be briefly described in the following three steps:
\begin{itemize}
  \item \emph{Key generation}.
  Assume that $\mbG$ and ${\mbG}_{T}$ are
  two (multiplication) cyclic groups of finite order $n$,
  and $e$ is a bilinear map.
  Let $g$ and $u$ be two random generators of $\mbG$,
  and $p, q$ be two big primes satisfying $n=pq$.
  Set $h = u^{q}$, then let
  $pk=(n, {\mbG} ,{\mbG}_{T}, e, g, h)$ and $sk = p$.

  \item \emph{Encryption}. A message $m$ can be encrypted to its ciphertext $c$ as follows
  \begin{center}
    $c = g^{m}h^{r} \in \mbG$,

    where $r$ is randomly picked in $\{0, 1, \dots, n-1\}$.
  \end{center}

  \item \emph{Decryption}. A ciphertext $c$ is decrypted as follows
  \begin{center}
    $c^{p} = {(g^{m}h^{r})}^{p} = g^{mp}u^{rpq}= {(g^{p})}^{m}$ $($mod $n)$.
  \end{center}
    Let $\hat{g}=g^{p}$.
    One needs to compute the discrete log of $c^{p}$ base $\hat{g}$ to recover $m$.
\end{itemize}

This scheme has the additive homomorphism over the encrypted data feature.
Given the ciphertext $E(a)$ and $E(b)$, we can get the result of $a+b$ by $E(a) \cdot E(b)$, i.e., $E(a+b) = E(a) \cdot E(b)$.
This feature allows us to calculate the sum of two numbers by their ciphertexts without decryption.

\section{Secure Phrase Search for Intelligent Processing of  Encrypted Data}\label{sec:main scheme}
This section presents the proposed privacy-preserving phrase search scheme over encrypted data.

\subsection{System Overview}\label{sec:mainidea}
\begin{figure}[t]
\begin{minipage}{0.49\textwidth}
  \centering
  \includegraphics[height=4.5cm]{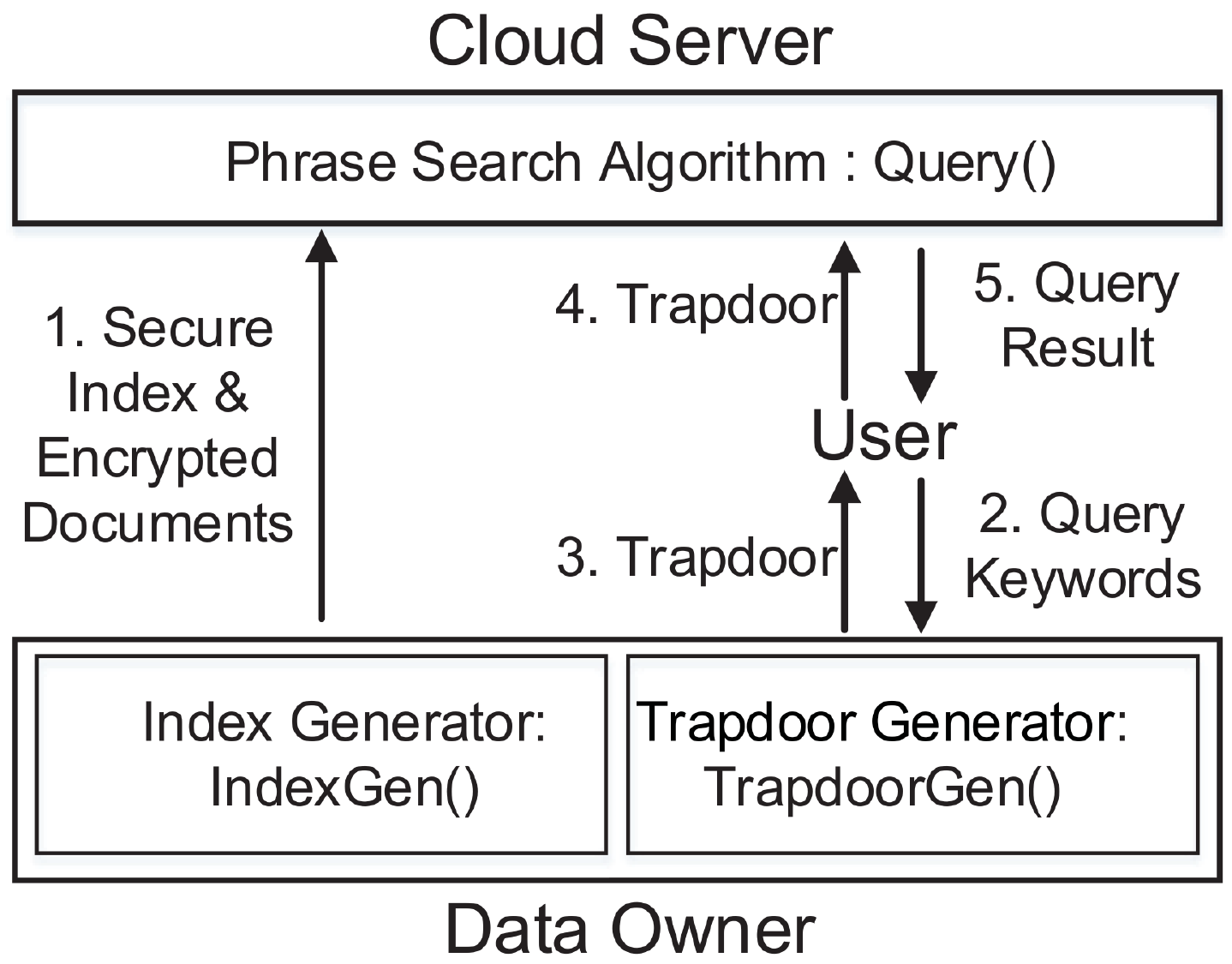}
\end{minipage}
\caption{The structure and workflow of the proposed scheme \texttt{P3}}\label{fig:System_Architecture}
\end{figure}

The structure and workflow of the proposed scheme, \texttt{P3}, are depicted in Fig. \ref{fig:System_Architecture},
which mainly consists of the following three modules:
\begin{itemize}
  \item \emph{Index Generator}, which is executed on the data owner side. It takes the documents as the input and outputs the corresponding secure index, as well as the encrypted documents.
  \item \emph{Trapdoor Generator}, which is also executed on the data owner side. Given a user's queried phrase, it generates the corresponding secure trapdoor and replies to the user.
  \item \emph{Phrase Search Algorithm}, which is executed on the cloud server side. Upon receiving a trapdoor from a user, it performs a phrase search procedure over the secure index and returns the search results.
\end{itemize}

\begin{figure}[t]
\begin{minipage}{0.49\textwidth}
  \centering
  \includegraphics[height=4.5cm]{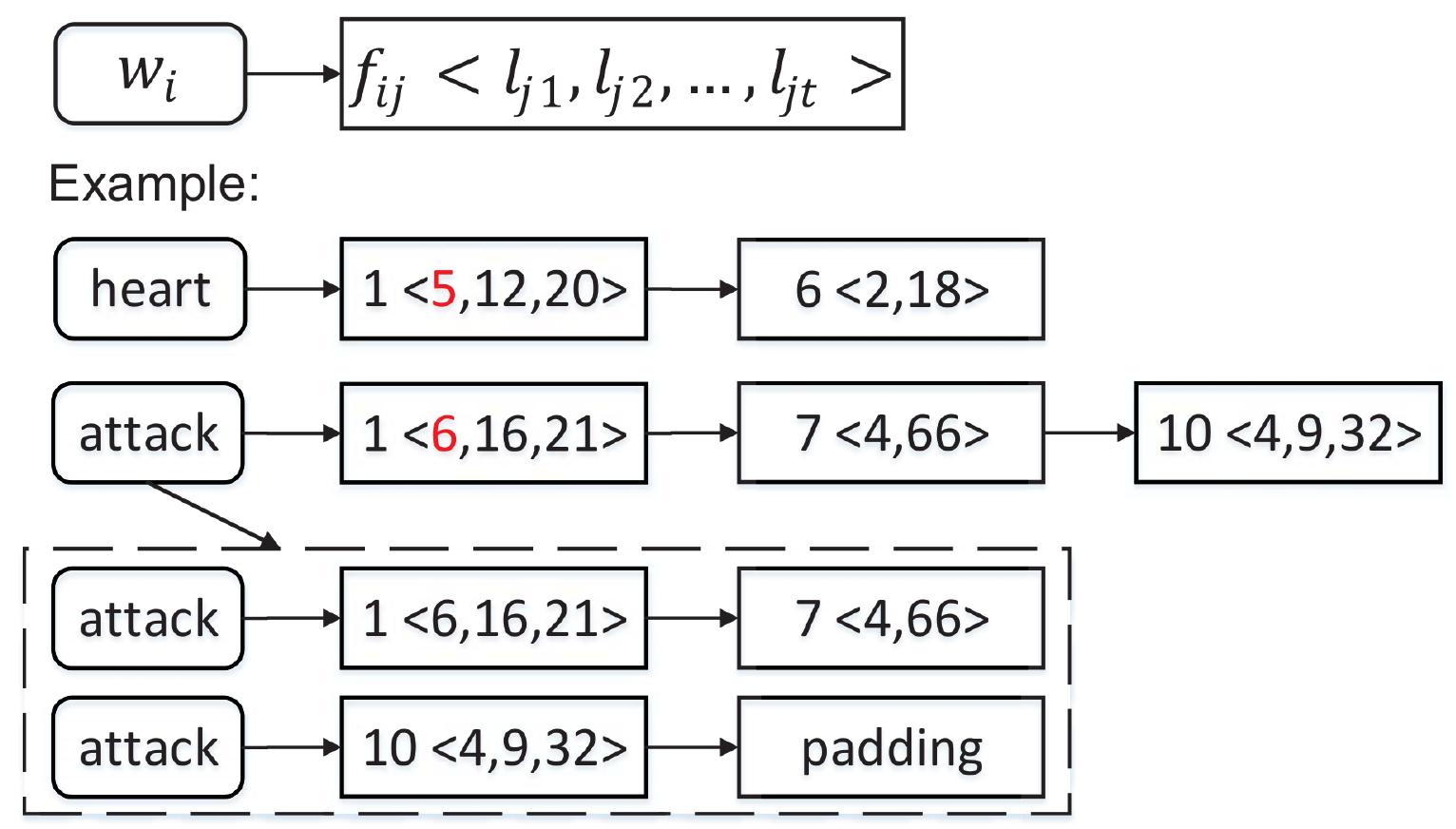}
\end{minipage}
\caption{An example of the inverted index. (Encryptions are not shown).}\label{fig:index_structure}
\end{figure}

In order to support phrase search,
we leverage the inverted index structure
and store the keyword locations along with the document identifier, as shown in Fig. \ref{fig:index_structure} (cf. Section \ref{subsec:notations} for explanations of notations). In the example illustrated in Fig. \ref{fig:index_structure}, there are two files containing the keyword \emph{heart}, namely Files 1 and 6. More precisely, the locations of \emph{heart} in File 1 are 5, 12, and 20, respectively.

The phrase search procedure can be described as follows.
When the cloud server receives the trapdoor for a specific phrase query from a user,
it first locates the inverted lists for the queried keywords,
and then finds the documents that contain all of the queried keywords.
After that, the cloud server identifies whether the locations of the keywords are consecutive and returns only the relevant documents that contain the exact phrase.
As shown in Fig. \ref{fig:index_structure}, File 1 should be returned if the user queries the phrase ``heart attack''.

It is easy to perform phrase search over plaintexts.
However, it is difficult for the server to determine whether or not the keywords occur in documents as a phrase,
given the encrypted location information of each pair of keywords.
To tackle this challenge,
we propose a series of designs based on the homomorphic encryption \cite{boneh2005evaluating} and bilinear groups.
We also utilize the widely used secure kNN method \cite{multi:keyword:ranked:search, multi:keyword:ranked:search:poly-func, Multi-Keyword-Fuzzy-Search-2014}
to achieve trapdoor unlinkability.

\subsection{Building Blocks}\label{sec:building_block}
As described in Section \ref{sec:mainidea},
we should ensure privacy in the index generation, the trapdoor generation, and the phrase search procedures.
We now introduce basic building blocks to achieve these goals.

1) \textbf{Keyword representation in the secure index and the trapdoor}.
We utilize a similar technique as in the literature \cite{multi:keyword:ranked:search:poly-func}
to achieve the goals of index privacy and trapdoor unlinkability.

Our design is based on the following observation.
Given a polynomial function $f(x)$ of degree $m$,
which is denoted by $f(x)=(x-t_{1})(x-t_{2}){\cdots}(x-t_{m})=a_{0}+a_{1}x+{\dots}+a_{m}x^{m}$,
we can extract the coefficients to form a vector $A=\{a_{0}, a_{1}, {\dots}, a_{m} \}$.
We can also construct another vector
$B={\{t^{0}, t^{1}, \dots , t^{m}\}}^{T}$,
where $t \in \{t_{1}, t_{2}, \dots , t_{m}\}$.
Note that $t^{m}$ represents $t$ to the power of $m$.
Since $t$ is a root of $f(x)$, we have $A^{T} \cdot B=0$.

Based on the above knowledge,
for any single keyword we construct two vectors, $A$ and $B$,
as its representations in the trapdoor and the index, respectively.
Then, we can know if a keyword in the trapdoor matches a keyword in the index by checking whether $A^{T} \cdot B=0$.
Hence, we now focus on constructing these two vectors for private-preserving matching.

To generate the encrypted keyword identifier $\widetilde{Z}_{w_{i}}$ for each keyword $w_{i} \in W$,
we utilize the secure kNN technique, as depicted in Algorithm \ref{Algorithm:Build_Keyword_Index}.
The algorithm includes two steps, where the first step is to create the vector $\widetilde{B}$ (line 1),
and the second step is to obtain the encrypted keyword identifier $\widetilde{Z}_{w_{i}}$ by splitting $\widetilde{B}$ randomly into two vectors ${\widetilde{B}}^{a}(i)$ and ${\widetilde{B}}^{b}(i)$ (lines 2-9).
According to the value of each element in $S$, ${\widetilde{B}}^{a}(i)$ and ${\widetilde{B}}^{b}(i)$ are assigned with different values.
We refer the readers to the literature \cite{multi:keyword:ranked:search, multi:keyword:ranked:search:poly-func, Multi-Keyword-Fuzzy-Search-2014} for the rationale of secure kNN.

\renewcommand{\algorithmicrequire}{\textbf{Input:}}
\renewcommand{\algorithmicensure}{\textbf{Output:}}

\begin{algorithm}[t]
  \small
  \caption{EncKeywordForIndex($\cdot$)}\label{Algorithm:Build_Keyword_Index}
  \begin{algorithmic}[1]
  \Require $\{w_{i}, K, S, M_{1}, M_{2}\}$,
  where $K$ is secret key of PRF $\pi$ and $S$, $M_{1}$,
  and $M_{2}$ are the secret keys of the secure kNN technique.
  Define $S(i)$ as the $i$-th bit in $S$.
  \Ensure The encrypted keyword identifier $\widetilde{Z}_{w_{i}}$ in the index.
  \State Construct a vector ${\widetilde{B}}={\{ \pi{(K, w_{i})}^{0}, \dots, \pi{(K, w_{i})}^{d-1}  \}}^{T}$,
  where $d$ is the length of $S$ and $\pi(\cdot)$ is a secure PRF primitive.
  \For {$i \gets 1$ \textbf{to} $d$}
       \If {$S(i) = 1$}
            \State Split $\widetilde{B}(i)$ randomly into ${\widetilde{B}}^{a}(i)$ and ${\widetilde{B}}^{b}(i)$ with ${\widetilde{B}}^{a}(i) + {\widetilde{B}}^{b}(i) = \widetilde{B}(i)$.
       \Else
            \State Set both ${\widetilde{B}}^{a}(i)$ and ${\widetilde{B}}^{b}(i)$ to $\widetilde{B}(i)$.
       \EndIf
  \EndFor
  \State Encrypt $\widetilde{B}$ as $M_{1}^{T}{\widetilde{B}}^{a}$ and $M_{2}^{T}{\widetilde{B}}^{b}$.
  \State Set $\widetilde{Z}_{w_{i}}=\{ M_{1}^{T}{\widetilde{B}}^{a}, M_{2}^{T}{\widetilde{B}}^{b} \}$.
  \State \textbf{return} $\widetilde{Z}_{w_{i}}$
  \end{algorithmic}
\end{algorithm}

To construct a secure trapdoor for a query $Q$,
we also utilize the secure kNN technique to construct the
encrypted keyword identifier $\widetilde{Y}_{w_{i}}$ for each keyword $w_{i} \in Q$,
as described in Algorithm \ref{Algorithm:Build_Trapdoor}.
It consists of two steps, where the first step (lines 1-4)
is to create the vector $\widetilde{A}$,
and the second step (lines 5-12) is to spilt $\widetilde{A}$ to
obtain the encrypted keyword identifier $\widetilde{Y}_{w_{i}}$.

\renewcommand{\algorithmicrequire}{\textbf{Input:}}
\renewcommand{\algorithmicensure}{\textbf{Output:}}

\begin{algorithm}[t]
  \small
  \caption{EncKeywordForTrapdoor($\cdot$)}\label{Algorithm:Build_Trapdoor}
  \begin{algorithmic}[1]
  \Require $\{w_{i}, K, S, M_{1}, M_{2}\}$,
  where $K$ is secret key of PRF $\pi$ and $S$, $M_{1}$,
  and $M_{2}$ are the secret keys of the secure kNN technique.
  Define $S(i)$ as the $i$-th bit in $S$.
  \Ensure The encrypted keyword identifier $\widetilde{Y}_{w_{i}}$ in the trapdoor.
  \State Construct a keyword vector $\Phi = \{ w_{i}, w'_{1}, \dots , w'_{d-2} \}$,
  where $d$ is the length of $S$ and $\{w'_{1}, \dots , w'_{d-2}\}$ are $d-2$ dummy keywords.
  \State Get a vector ${\widetilde{\Phi}}={\{ \pi{(K, w_{i})}, \pi{(K, w'_{1})}, \dots, \pi{(K, w'_{d-2})} \}}$,
  where $d$ is the length of $S$ and $\pi(\cdot)$ is a secure PRF primitive.
  \State Construct a polynomial function of degree $d-1$ as
  $f(x)=(x-\pi{(K, w_{i})}) \times (x-\pi{(K, w'_{1})}) \times \dots \times (x-\pi{(K, w'_{d-2})})=a_{0}+a_{1}x+ \dots + a_{d-1}x^{d-1}$.
  \State Extract the coefficients of $f(x)$ to form the query vector $\widetilde{A}={\{ a_{0}, a_{1}, \dots , a_{d-1}\}}^{T}$.
  \For {$i \gets 1$ \textbf{to} $d$}
       \If {$S(i) = 0$}
            \State Split $\widetilde{A}(i)$ randomly into ${\widetilde{A}}^{a}(i)$ and ${\widetilde{A}}^{b}(i)$, where ${\widetilde{A}}^{a}(i) + {\widetilde{A}}^{b}(i) = \widetilde{A}(i)$.
       \Else
            \State Set both ${\widetilde{A}}^{a}(i)$ and ${\widetilde{A}}^{b}(i)$ to $\widetilde{A}(i)$.
       \EndIf
  \EndFor
  \State Encrypt $\widetilde{A}$ as $M_{1}^{-1}{\widetilde{A}}^{a}$ and $M_{2}^{-1}{\widetilde{A}}^{b}$.
  \State Set $\widetilde{Y}_{w_{i}}=\{ M_{1}^{-1}{\widetilde{A}}^{a}, M_{2}^{-1}{\widetilde{A}}^{b} \}$.
  \State \textbf{return} $\widetilde{Y}_{w_{i}}$
  \end{algorithmic}
\end{algorithm}

Based on the above constructions, given an encrypted keyword identifier $\widetilde{Y}_{w_{i}}$ in a trapdoor,
the cloud server can locate an inverted list with an encrypted keyword identifier $\widetilde{Z}_{w_{i}}$, by checking whether
$\widetilde{Y}_{w_{i}}^{T} \cdot \widetilde{Z}_{w_{i}}=0$.

The correctness of this construction is illustrated by Eq. \eqref{eq:kNN}.
  \begin{align}\label{eq:kNN}
    \footnotesize
    \begin{split}
  {\widetilde{Y}_{w_{i}}^{T} \cdot \widetilde{Z}_{w_{i}}}= &{ \{ M_{1}^{-1}{\widetilde{A}}^{a}, M_{2}^{-1}{\widetilde{A}}^{b} \} }^{T} \cdot \{ M_{1}^{T}{\widetilde{B}}^{a}, M_{2}^{T}{\widetilde{B}}^{b} \} \\
  =&{({\widetilde{A}}^{a})}^{T}{(M_{1}^{-1})}^{T}M_{1}^{T}{\widetilde{B}}^{a}+{({\widetilde{A}}^{b})}^{T}{(M_{2}^{-1})}^{T}M_{2}^{T}{\widetilde{B}}^{b}\\ =&{({\widetilde{A}}^{a})}^{T}{\widetilde{B}}^{a}+{({\widetilde{A}}^{b})}^{T}{\widetilde{B}}^{b}
  ={\widetilde{A}}^{T} \cdot \widetilde{B}
    \end{split}
  \end{align}

The secure kNN method is vulnerable to linear analysis,
and this means that the cloud server may launch the linear analysis on a large number of pairs of keyword identifiers between the secure index and the trapdoors.
To address this limitation, we adopt dummy keywords in the procedure of trapdoor generation (lines 1-3 in Aigorithm \ref{Algorithm:Build_Trapdoor}).
Therefore, for the same keyword over multiple queries,
we can obtain a different coefficient vector $\widetilde{A}$ (line 4 in Algorithm \ref{Algorithm:Build_Trapdoor}).
Furthermore,
due to the property of the secure KNN technique,
we can perform various splittings over a coefficient vector $\widetilde{A}$.
Hence, our construction is secure against linear analysis.

2) {\textbf{Phrase recognition}.
To protect the keyword location privacy,
we encrypt the keyword location through the homomorphic encryption scheme introduced in Section \ref{subsec:Preliminaries}.

Note that in our scheme, we only publish $(n, \mbG , {\mbG}_{T}, e)$ to the cloud server as the public key.
Assume that $a$ and $b$ represent locations of two different keywords in a same document.
Without loss of generality, we also assume that $a<b$.
If these two keywords are consecutive, we have $a-b+1=0$, i.e.,
$b-a=1$.
To determine the relationship between $a$ and $b$ on the basis of their ciphertexts $g^{a}h^{r_{1}}$ and $g^{b}h^{r_{2}}$,
the cloud server sets $x=a-b+1$ and transforms this problem to an equivalent problem of determining whether $x$ is the ciphertext of $0$, as shown in Eq. \eqref{eq:zero},
\begin{align}\label{eq:zero}
\begin{split}
    E(x) & = E(a-b+1) \\
         & = g^{a}h^{r_{1}} \cdot {(g^{b}h^{r_{2}})}^{-1} \cdot g^{1}h^{r_{3}} = g^{x}h^{r}
    \end{split}
\end{align}
where $g^{1}h^{r_{3}}$ represents the ciphertext of $1$.

Then, the cloud server further determines the relationship between $a$ and $b$ depending on the result of Eq. \eqref{eq:one},
\begin{align}\label{eq:one}
    e(E(x), {\lambda}^{p}) = e(g^{x}h^{r}, {\lambda}^{p})
\end{align}
where $\lambda \in \mbG$, $p$ is the private key, and ${\lambda}^{p}$ is the dispersal factor that cannot be an identity of $\mbG$.

Until now, the cloud server has known $g^{x}h^{r}$ and ${\lambda}^{p}$.
To eliminate the random value $r$, it then computes $e(g^{x}h^{r},{\lambda}^{p})$ by bilinear maps.
Note that $a$ and $b$ represent consecutive locations if and only if the result of Eq. \eqref{eq:one} is equal to $1$,
as $e(g^{x}h^{r}, {\lambda}^{p}) = e(g^{0}h^{r},{\lambda}^{p})=e(h^{r},{\lambda}^{p})=e{(h,\lambda)}^{rp}=e(h^{rp}, \lambda)=e(1,\lambda)=1$.

The idea of such a design comes from the fact that we can eliminate the existence of the random value $r$ for ${(h^{r})}^{p} = {u}^{rpq} = {u}^{rn} = 1$ $($mod $n)$.
However, since the phrase recognition procedure is performed by the cloud server, a user cannot send $p$ to the cloud server directly.
Therefore, the user randomly picks an element $\lambda \in \mbG$ and sends ${\lambda}^{p}$ to the cloud server.
Since $\lambda$ and $p$ are both secret, the cloud server cannot infer $p$ from ${\lambda}^{p}$.

Now we briefly discuss the construction of the phrase recognition process.
First, at a high level,
we want to protect the keyword location information, rather than the keyword location relationship in the phrase search.
This is because revealing the keyword location relationship is inevitable to perform phrase recognition.
Second,
the recognition method can determine an arbitrary interval for two integers.
In other words, if we want to know whether the interval between two locations $a$ and $b$ is $d$,
we can just send $g^{d}h^{r}$ to the cloud server,
where $r$ is a random number.
In addition, the ciphertexts for the same $d$ over multiple queries are different.
This property can prevent the cloud server from inferring the interval $d$, because the cloud server cannot know the real value of $d$ even if it learns that $a$ and $b$ satisfy a certain relationship.

Note that this application scenario is different from the well-known secure multi-party computation (i.e., SMC).
In the setting of SMC,
set of parties with private inputs wish to compute a function of their inputs while revealing nothing but the result of the function,
which is used for many practical applications such as exchange markets.
SMC is a collaborative computing problem that solves the privacy preserving problem among a group of mutually untrusted participants.
Thus the SMC schemes are fully secure,
they protect the location relationship between keywords against the cloud server.
As a result, the phrase recognition procedure can only be performed on either the user side or the data owner side,
which sacrifices the main benefit of offloading computation to cloud servers.
Therefore, we make a compromise that revealing the relationship between keyword locations for better efficiency.

3) \textbf{Division and padding of inverted list}.
To protect the keyword privacy, it is necessary to hide its appearance frequency in each document.
%
We divide each inverted list to make it contain $\eta$ documents.
Then, if the length of an (original or divided) inverted list is smaller than $\eta$, we perform a padding for the remaining entries.
In the example shown in Fig. \ref{fig:index_structure},
we choose $\eta = 2$ and divide ``attack'' into two inverted lists, where the second list has a padding entry.
More precisely, each entity that we pad consists of an invalid document identifier and some random numbers as fake keyword locations.

In order to distinguish these invalid document identifiers
from the valid ones,
we use a counter that is initialized as $-1$
and gradually decrease it by 1 for each padded document identifier.
Due to the encryption of the invalid and valid document identifiers,
the cloud server cannot tell which document identifer is invalid.

Since we utilize the probabilistic encryption, a same keyword $w_i$ will have different ciphertexts (i.e., the encrypted keyword identifier $\widetilde{Z}_{w_{i}}$).
Therefore, from the perspective of the cloud server, it seems that each inverted list corresponds to a unique keyword.
In the performance evaluation, we select $\eta$ as the frequency median of all the keywords in the document set. We leave the exploration of optimal $\eta$ to the future work.

\subsection{Scheme Details}
This section describes the privacy-preserving phrase search scheme in detail,
which consists of four components.

$\bullet$ \textbf{KeyGen$(\tau, d)$:}
Given the security parameters $\tau$ and $d$, the data owner generates the master key and the public key by taking the following steps:

1) Generate two random $\tau$-bit big primes $p$ and $q$, and set $n=p*q$.
Construct the bilinear groups $\mbG$ and ${\mbG}_{T}$ and the bilinear map $e$ using the method introduced in the literature \cite{boneh2005evaluating}.
Then, pick two random generators, $g$ and $u$, from $\mbG$, and set $h=u^{q}$.
Note that $h$ is a random generator of the subgroup of $\mbG$
of order $p$.

2) Randomly generate a $d$-bit binary string $S$ and
two $d \times d$ invertible matrices $M_{1}$ and $M_{2}$.
Let $S(i)$ be the $i$-th bit of $S$.

3) Let $\pi$ be a secure pseudorandom function (PRF) primitive and generate a $\tau$-bit secret key $K$.

4) Let $\nu$ be a secure pseudorandom permutation (PRP) primitive and generate a $\tau$-bit secret key $U$.

The data owner keeps the tuple $(p,g, h, K, U, S, M_{1}, M_{2})$ as the master key (i.e., $Mk$) and the tuple $(n,\mbG, {\mbG}_{T}, e)$ as the public key (i.e., $pk$),
which is published to the cloud server.

$\bullet$ \textbf{IndexGen$(Mk,\Gamma)$:}
The data owner builds the secure inverted index in the following steps:

1) Extract a distinct keyword collection $W$ of size $\mu$ from the document collection $\Gamma$.
For each keyword $w_{i}\in{W} (1\le{i}\le{\mu})$, build the inverted list $\mbI_{w_{i}}$ as described in Fig. \ref{fig:index_structure}, which consists of the identifiers of documents that contain keyword $w_{i}$ along with all the keyword locations, i.e., ${\mbI}_{w_{i}}=(w_{i}, \Omega_{i1}, \Omega_{i2}, \dots, \Omega_{ik})$, where $\Omega_{ij} = (f_{ij}, \Lambda_{ij})$ and $\Lambda_{ij} = \langle l_{j1}, l_{j2}, \dots, l_{jt} \rangle$, ${1} \le j \le {k}$, $1\le{r}\le{t}$.
Set the inverted index $\mbI = \{ \mbI_{w_{1}}, \mbI_{w_{2}}, \dots , \mbI_{w_{\mu}} \}$.

2) For each $\mbI_{w_{i}} \in {\mbI}$,
encrypt the document identifier by $\nu(U, f_{ij})$
and encrypt the keyword locations.

More precisely, for each location $l_{jy} \in {\Lambda_{ij}}$, pick a random number $r_{jy} \in \{ 0,1, \dots , n-1 \}$; then, we have the encrypted location $c_{jy}$ as described by Eq. \eqref{eq:loc_encryption}.
\begin{align}\label{eq:loc_encryption}
    c_{jy} = {g^{l_{jy}} {h ^{r_{jy}}}}
\end{align}

To hide keyword frequencies, we should guarantee that different keywords have the same frequency.
Hence, the data owner should further process $\mbI_{w_{i}}$ via division and padding.

If $\lfloor |\mbI_{w_{i}}|$ \emph{mod} $\eta \rceil =0$,
divide $\mbI_{w_{i}}$ into $|\mbI_{w_{i}}|/\eta$ individual inverted lists,
which are defined as $\{ \mbI_{w_{i1}}, \mbI_{w_{i2}}, \dots , \mbI_{w_{it}} \}$, $1 \le t \le |\mbI_{w_{i}}|/\eta$.
While if $\lfloor |\mbI_{w_{i}}|$ \emph{mod} $\eta \rceil \ne 0$,
divide $\mbI_{w_{i}}$ into $1 + |\mbI_{w_{i}}|/\eta$ individual inverted lists, which are defined as $\{ \mbI_{w_{i1}}, \mbI_{w_{i2}}, \dots , \mbI_{w_{it}} \}$, $1 \le t \le 1+|\mbI_{w_{i}}|/\eta$.
For $\mbI_{w_{it}}$ where $t=1+|\mbI_{w_{i}}|/\eta$, the data owner pads some random dummy document identifiers and binary strings of length $|c_{jy}|$
to make sure that the keyword document frequency is $\eta$.

3) For each inverted list $\mbI_{w_{it}}$ of the keyword $w_{it}$,
encrypt the keyword $w_{it}$ to obtain the encrypted keyword identifier $\widetilde{Z}_{w_{it}}$ using Algorithm \ref{Algorithm:Build_Keyword_Index}.
Then, update $w_{it}$ with $\widetilde{Z}_{w_{it}}$.
We now get the secure inverted index $\widehat{\mbI}=\{\{ \widehat{\mbI}_{w_{it}} \}\}$, where $1 \le i \le \mu$.

$\bullet$ \textbf{TrapdoorGen$(Mk,Q)$:}
Given a query $Q$, a user can retrieve the corresponding trapdoor from the data owner, which takes the following steps:

1) For each $w_{j} \in Q$, $1 \le j \le |Q|$,
generate the encrypted query keyword identifier $\widetilde{Y}_{w_{j}}$ using Algorithm \ref{Algorithm:Build_Trapdoor}.

2) We assume that the search distance $\beta$ is 1.
Pick a random number $r \in [0, n-1]$, and then compute the ciphertext of 1:
\begin{align}\label{eq:mbC}
    \mbC = g^{1} {h^{r}}
\end{align}

3) Randomly pick an element $\lambda \in \mbG$,
and then compute the dispersal factor, $\psi = \lambda ^ {p}$,
where ${\lambda}^{p}$ is not an identity of $\mbG$.

The trapdoor $\mbT_{Q}=\{ \{ \widetilde{Y}_{w_{j}} \}, \mbC, \psi \}$, where $1 \le j \le |Q|$.

$\bullet$ \textbf{Query$(\widehat{\mbI},{\mbT}_{Q})$:}
Once the cloud server receives the trapdoor from the user, the cloud server first locates the inverted lists corresponding to the queried keywords, by checking whether ${\widetilde{Y}_{w_{j}}}^{T} \cdot \widetilde{Z}_{w_{it}}$ is equal to 0.
As described in Section \ref{sec:building_block},
an equality indicates a match of the queried keyword and the inverted list.
We assume that the corresponding inverted lists are $\widehat{\mbI}_{Q}=\{\widehat{\mbI}_{w_{i}}\}$, where $1 \le i \le k$, i.e., $k = |\widehat{\mbI}_{Q}|$.
Then, the server identifies the documents containing the exact queried phrase by determining the positional relationship of the keywords using Eqs. \eqref{eq:zero} and \eqref{eq:one}.
Finally, the server replies to the user with the search results, i.e., the corresponding encrypted documents that contain the queried phrase.



\section{Security Analysis}\label{sec:security}
This section presents the security analysis under the \emph{known ciphertext model} and the \emph{known background model}.
We adopt the security definitions in the literature \cite{Curtmola:2006:SSE:1180405.1180417}.
\begin{itemize}
  \item \emph{History}. Let $\Gamma$ be a file set and $\mbI$ be the index built from $\Gamma$.
      A \emph{history} over $\Gamma$ is a tuple $H=(\Gamma , \mbI, w)$, where $w$ is a phrase containing $k$ keywords $w=(w_{1}, w_{2}, \dots, w_{k})$.

  \item \emph{View}, denoted by $V(H)$, is the encrypted form of $H$ under a certain secret key $sk$.
  In general, a $V(H)$ consists of
  the encrypted documents $Enc_{sk}(\Gamma)$,
  the secure index $Enc_{sk}(\mbI(\Gamma))$,
  and the secure trapdoor $Enc_{sk}(w)$.
  Note that the cloud server can only know the views.

  \item \emph{Trace}.
  The trace of history, which is denoted by $Tr(H)$, consists of exactly the information we are willing to leak about the history and nothing else.
  More precisely, it should be the access patterns and the search results induced by $H$.
  The trace induced by a \emph{history} $H=(\Gamma , \mbI, w)$, is a sequence $Tr(H) = Tr(w) = \{ {R_{w}, ({\delta}_{i})}_{w \subset {\delta}_{i}}, 1 \le i \le |\Gamma| \}$,
  where $w$ should occur in the document ${\delta}_{i}$
  as a phrase,
  and $R_{w}$ indicates whether these keywords constitute a phrase in the documents.
\end{itemize}

\textbf{Theorem 1.} \emph{Our phrase search scheme is secure under the known ciphertext model.}

Intuitively, given two histories with the same trace,
if the cloud server cannot distinguish which one is generated by a simulator,
we can say that it cannot learn additional information about the secure index or the encrypted documents,
except for the access patterns and search results.

\begin{proof}
Assume that $S$ is a simulator that can simulate a view $V^{'}$ indistinguishable from the view obtained by the cloud server.
To achieve this, we construct the simulator as follows:

\begin{itemize}
  \item $S$ selects a random ${\delta}_{i}^{'} \in {\{ 0, 1 \}}^{|{\delta}_{i}|}, {\delta}_{i} \in \Gamma, 1 \le i \le |\Gamma|$, and then outputs
  ${\Gamma}^{'} = \{ {\delta}_{i}^{'}, 1 \le i \le |{\Gamma}^{'}| \}$.

  \item $S$ first generates two random $\tau$-bit big primes $p'$ and $q'$ to obtain $n' = p'*q'$, and constructs the bilinear groups $\mbG'$ and $\mbG_{T}^{'}$.
  Then, $S$ selects two random generators $g'$ and $u'$ from $\mbG'$ and obtains $h' = {u'}^{q'}$.
  Finally, $S$ randomly picks a $d$-bit binary string $S'$,
  two $d \times d$ invertible matrices $M_{1}^{'}, M_{2}^{'}$,
  a secure hash function $\pi(\cdot)$ with a secret key $K'$,
  and a secure pseudo-random permutation (PRP) primitive $\nu$ with the secret key $U'$.
  Let ${sk}' = \{ p', g', h', K', U', S', M_{1}^{'}, M_{2}^{'} \}$.

  \item $S$ generates $\mbI'({\Gamma}^{'})$ with the same dictionary $W$ as $\Gamma$.
   For each $w_{i} \in W$, $S$ takes the following steps:

   1) $S$ picks a random binary string as the inverted list $\mbI'_{w_{i}}$, which has the same length as the actual inverted list $\mbI_{w_{i}}$.
   Ensure that if $w_{i} \in W$ and $w_{i} \subset {\delta}_{i}, 1 \le i \le |\Gamma|$, the inverted list $\mbI'_{w_{i}}$ should contain the identifier $\nu(U', id({\delta}_{i}))$ of ${\delta}_{i}$.
   Meanwhile, if $w$ occurs in ${\delta}_{i}$ as a phrase, we should also ensure that $w$ occurs in ${\delta}_{i}^{'}$ as a phrase.

  2) $S$ gets ${\widetilde{B'}}={\{ \pi{(K', w_{i})}^{0}, 
  \dots, \pi{(K', w_{i})}^{d-1}\}}^{T}$ and computes $Enc_{sk'}(\widetilde{B'})$.
  Finally, $S$ obtains $Enc_{sk'}(\mbI'({\Gamma}'))$.

  \item $S$ constructs the query $w'$ and the corresponding trapdoor as follows.
  For each $w_{i} \in w$, $S$ constructs the encrypted keyword identifier $\widetilde{Y}_{w_{j}}$ by Algorithm \ref{Algorithm:Build_Trapdoor}.
  Then $S$ sets $Enc_{sk'}(w') = \{ \{ \widetilde{Y}_{w_{j}} \}, Enc_{sk'}(1), {{\lambda}'}^{p'} \}$ as the trapdoor, where ${\lambda}'$ is a random element of $\mbG'$ and $1 \le j \le |w'|$.

  \item Finally, $S$ outputs the view $V' = ( {\Gamma}', Enc_{sk'}(I'({\Gamma}')), Enc_{sk'}(w') )$.
\end{itemize}

The correctness of the construction is easy to demonstrate,
as the secure index $Enc_{sk'}(I'({\Gamma}'))$ and
the trapdoor $Enc_{sk'}(w')$ generate the same trace as
the one obtained by the cloud server.
Hence, we can claim that for any probabilistic polynomial-time (P.P.T.) adversary, $V'$ cannot be distinguished from $V(H)$.
Furthermore, no P.P.T. adversary can distinguish the ${\Gamma}'$ from $Enc_{sk}(\Gamma)$ for the semantic security of the symmetric encryption.
The indistinguishability of the index and trapdoors are guaranteed and enhanced together by
the indistinguishability of the secure kNN technique,
the random number introduced in the splitting process, and the use of probabilistic encryption.
\end{proof}

\textbf{Theorem 2.} \emph{Our phrase search scheme is secure under the known background model.}

Intuitively, given a view generated by the simulator,
if the cloud server, who has several pairs of queried phrases and trapdoors, cannot distinguish it from the view he owns,
we can say that the proposed phrase search scheme is secure under the known background model.

\begin{proof}
Based on the above construction, we can claim that no P.P.T. adversary can distinguish the view $V'$ from $V(H)$ with
a certain number of pairs of keywords and trapdoors.
Particularly, no P.P.T. adversary can distinguish the ${\Gamma}'$ from $Enc_{sk}(\Gamma)$ for the semantic security of the symmetric encryption.
Due to the usage of the dummy keywords and the probabilistic encryption,
the same queries will have different trapdoors.
Therefore, the P.P.T. adversary cannot launch the linear analysis using the pairs of queried phrases and trapdoors.
Thus, the indistinguishability of indices and trapdoors are guaranteed.
\end{proof}

\section{Performance Evaluation}\label{sec:evaluation}
In this section, we evaluate the performance of \texttt{P3} through extensive experiments using real-world datasets.

\subsection{Experiment Setup}

\textbf{Testbed}.
To simulate the cloud-based service environment,
we use an Aliyun server instance\footnote{https://www.aliyun.com/} as the cloud server,
which is equipped with an Intel Xeon processor at 2.60 GHz and 8 GB RAM.

\textbf{Dataset}.
We use a collection of the RFCs (Requests for Comments \cite{RFC}) as the real-world dataset for evaluation.
Each file contains a large number of technical phrases, e.g., \emph{error detection}.
We randomly pick up 2500 files from the publicly available RFCs.
For each file in the dataset, we build a full-text index, which is the same as the one commonly used by modern search engines.

\begin{figure*}[t]
\begin{minipage}{0.5\textwidth}
    \centering
    \includegraphics[height=4.5cm]{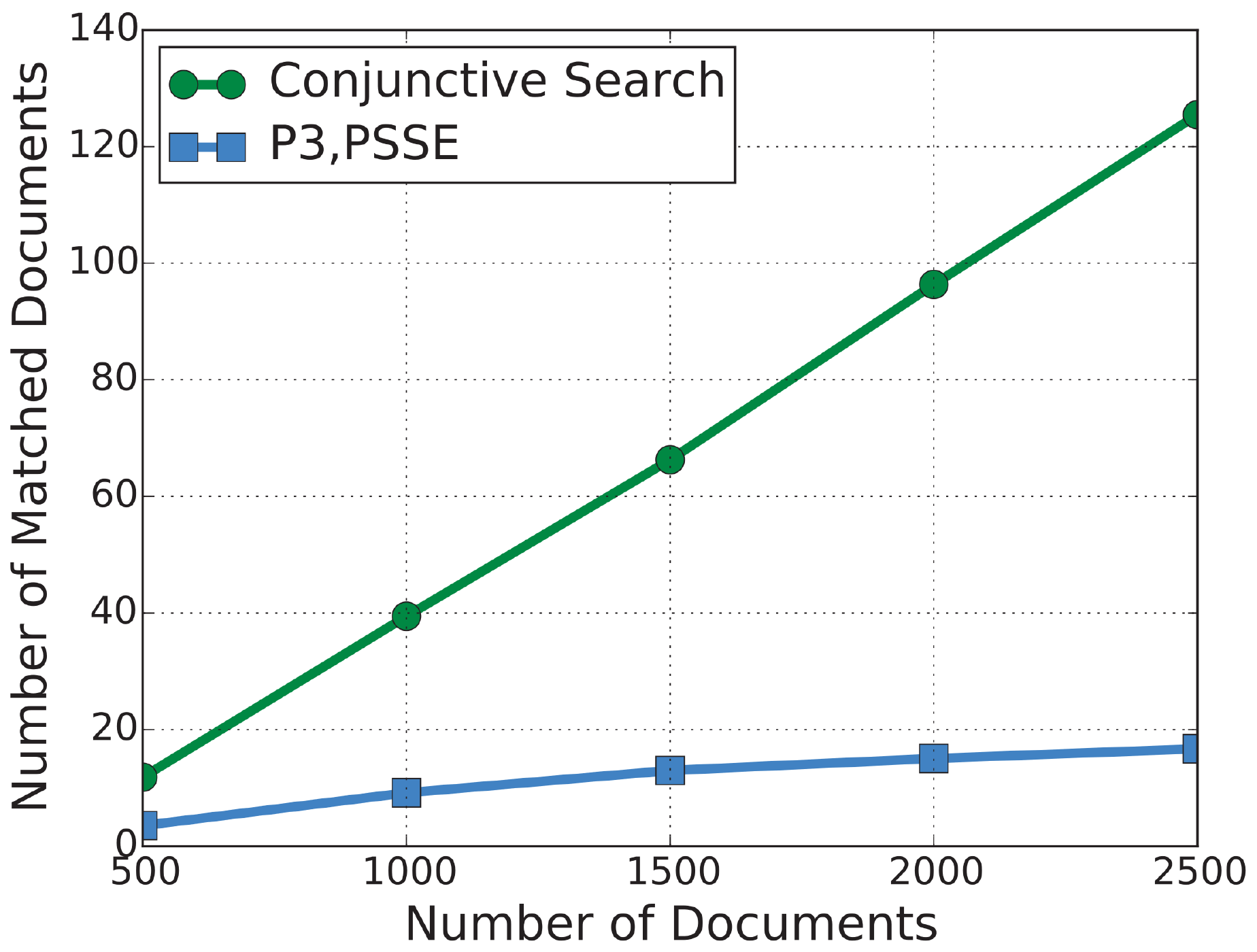}\\
    \scriptsize{(a) Number of documents in search results ($|Q|=2$)}
\end{minipage}
\quad
\begin{minipage}{0.5\textwidth}
    \centering
    \includegraphics[height=4.5cm]{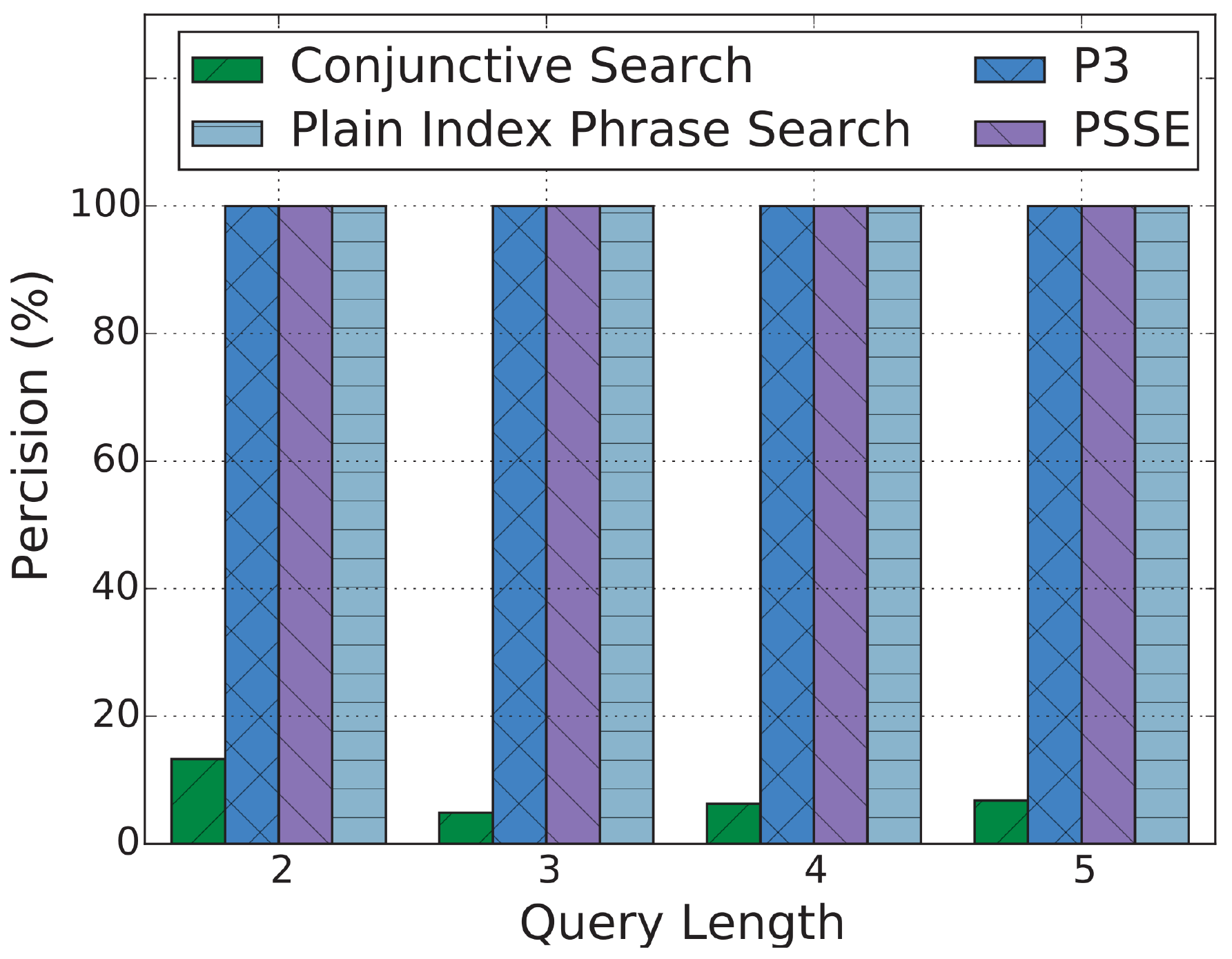}\\
    \scriptsize{(b) Precision with different query lengths ($m=2500$)}
\end{minipage}
\caption{Search accuracy with varying document sets and query lengths}\label{fig:subfig_accuracy}
\end{figure*}

\textbf{Methods to compare}.
We compare \texttt{P3} with a representative phrase search solution \cite{seaond:phrase:search}
and the traditional multi-keyword conjunctive search scheme,
which are referred to as \emph{PSSE} and \emph{conjunctive search}, respectively.
Since an implementation of PSSE is not given in the literature \cite{seaond:phrase:search},
we implement it using Java.
The conjunctive search scheme can be implemented simply by ignoring the phrase recognition procedure in \texttt{P3}.
Although it is not an exact implementation of an existing solution in the literature, it can still help us to understand the differences of the results returned by the conjunctive multi-keyword search and the phrase search.
We use a 128-bit security parameter in all the three methods.

The threshold parameter $\eta$ is set to be 32, which is
as the frequency median of all keywords in the document set.
We denote $|Q|$ as the phrase length
(i.e., the number of keywords in the phrase)
and $m$ as the number of documents.

\textbf{Query sets}.
We generate the querying phrases by randomly choosing phrases with semantics from the file set, e.g., sophisticated terminals, interrupt characters, shared memory, etc.
We use the same query length setting as existing studies \cite{2014_Phrase_query_optimization_on_inverted_indexes},
where $|Q|$ takes the concrete values of 2, 3, 4, and 5.

\subsection{Search Accuracy}\label{sec:search_accuracy}

We adopt a definition of the search accuracy widely used in the literature \cite{Multi-Keyword-Fuzzy-Search-2014}.
Given a phrase query, the search accuracy $\mathcal{P}$ is calculated as $\mathcal{P} = t_{p}/(f_{p}+t_{p})$,
where $t_{p}$ and $f_{p}$ are the numbers of relevant (i.e., containing the exact phrase) and irrelevant (i.e., containing all the keywords rather than the exact phrase) documents in the search results.

We first fix $|Q|=2$ and explore the numbers of matched documents for each method with varying scales of the document set, as shown in Fig. \ref{fig:subfig_accuracy}(a).
Compared with the conjunctive search scheme,
\texttt{P3} and PSSE can remarkably reduce the number of matched documents.

The precision with respect to different query lengths for each method is depicted in Fig. \ref{fig:subfig_accuracy}(b).
Here, the plain index phrase search scheme serves as the baseline of the precision.
We can see that the precisions of \texttt{P3} and PSSE are $100\%$ in all the cases,
whereas those of the conjunctive search scheme are less than $20\%$ in all the cases.

\subsection{Search Efficiency}
\textbf{Index Construction.}
The index construction process is a one-time, offline computation.
The time and storage overheads of the index construction
are depicted in Table \ref{tab:index}.
Clearly, the overheads increase when the document set gets larger.
For the same document set,
the index size of PSSE is much larger than that of \texttt{P3}.
As to the index construction time,
\texttt{P3} requires slightly more time than PSSE,
which is primarily caused by the encryption operations of the keyword locations.

\begin{table}[t]
\small
\renewcommand{\arraystretch}{1}
\caption{Summary of Index Construction Overheads} \label{tab:index} \centering
\centering
\footnotesize{
\begin{tabular}{|C{0.9cm}|C{0.9cm}|C{0.7cm}|C{0.7cm}|C{0.7cm}|C{0.7cm}|C{0.7cm}|}
\hline
\multicolumn{2}{|c|}{\multirow{2}{*}{Metrics / Methods}} &
\multicolumn{5}{c|}{Number of Documents} \\
\cline{3-7}
\multicolumn{2}{|c|}{} & 500 & 1000 & 1500 & 2000 & 2500 \\ \hline
\multirow{2}{*}{\tabincell{c}{ Time \\ (h) }} &
PSSE & 0.23 & 2.9 & 5.4 & 8.3 & 11.8 \\
\cline{2-7}
 & \texttt{P3} & 0.64 & 3.5 & 7.1 & 11.2& 15.1 \\ \hline

\multirow{2}{*}{\tabincell{c}{Volume \\ (GB)}}  &
PSSE & 1.48 & 19.9 & 35.9 &57.5 & 78.3 \\
\cline{2-7}
 & \texttt{P3} & 0.08 & 0.31 & 0.56 & 0.87 & 1.15 \\ \hline
\end{tabular}}
\end{table}

\textbf{Trapdoor Generation.}
The trapdoor generation time for each method with different query lengths is depicted in Fig. \ref{fig:subfig_trapdoor:c}.
$\texttt{P3}$ has a higher time cost of trapdoor generation than the conjunctive search scheme, because it needs extra operations (e.g., generating dispersal factor) to generate additional information for phrase judgement.
Compared with PSSE,
\texttt{P3} can reduce the time cost, especially when the query length is less than 8.
This is because PSSE needs two rounds of interaction between the user and the cloud server, and during the second interaction, it needs to generate a trapdoor for each document that was returned in the first interaction.
As the query length increases,
the number of documents returned in the first interaction could drop, which leads to a fall of the trapdoor generation time for PSSE.

\begin{figure*}
\begin{minipage}[t]{0.32\textwidth}
    \centering
    \includegraphics[height=4.5cm]{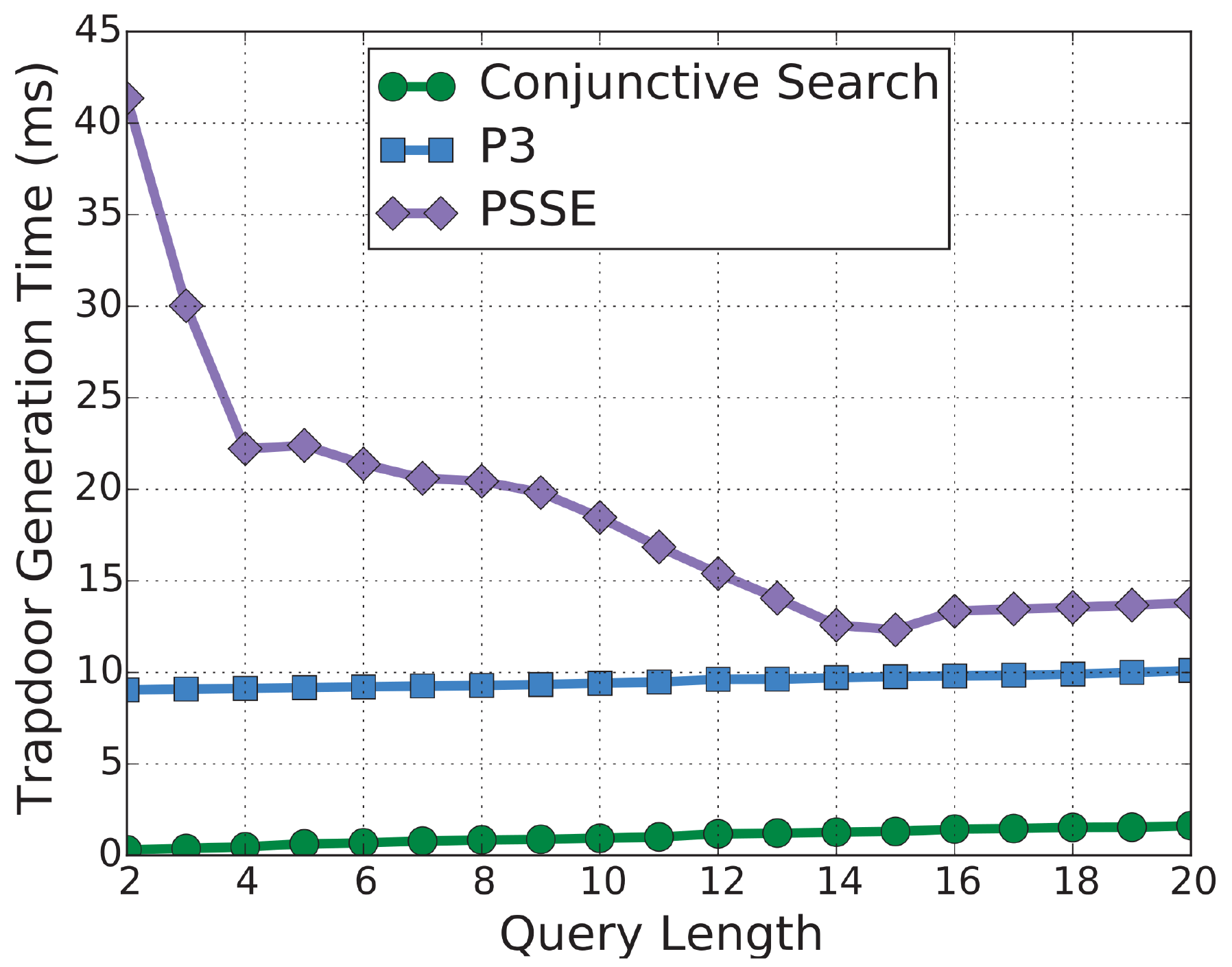}
    \caption{Trapdoor generation time for different query lengths ($m=2500$).}
    \label{fig:subfig_trapdoor:c}
\end{minipage}%
\quad
\begin{minipage}[t]{0.32\textwidth}
    \centering
    \includegraphics[height=4.5cm]{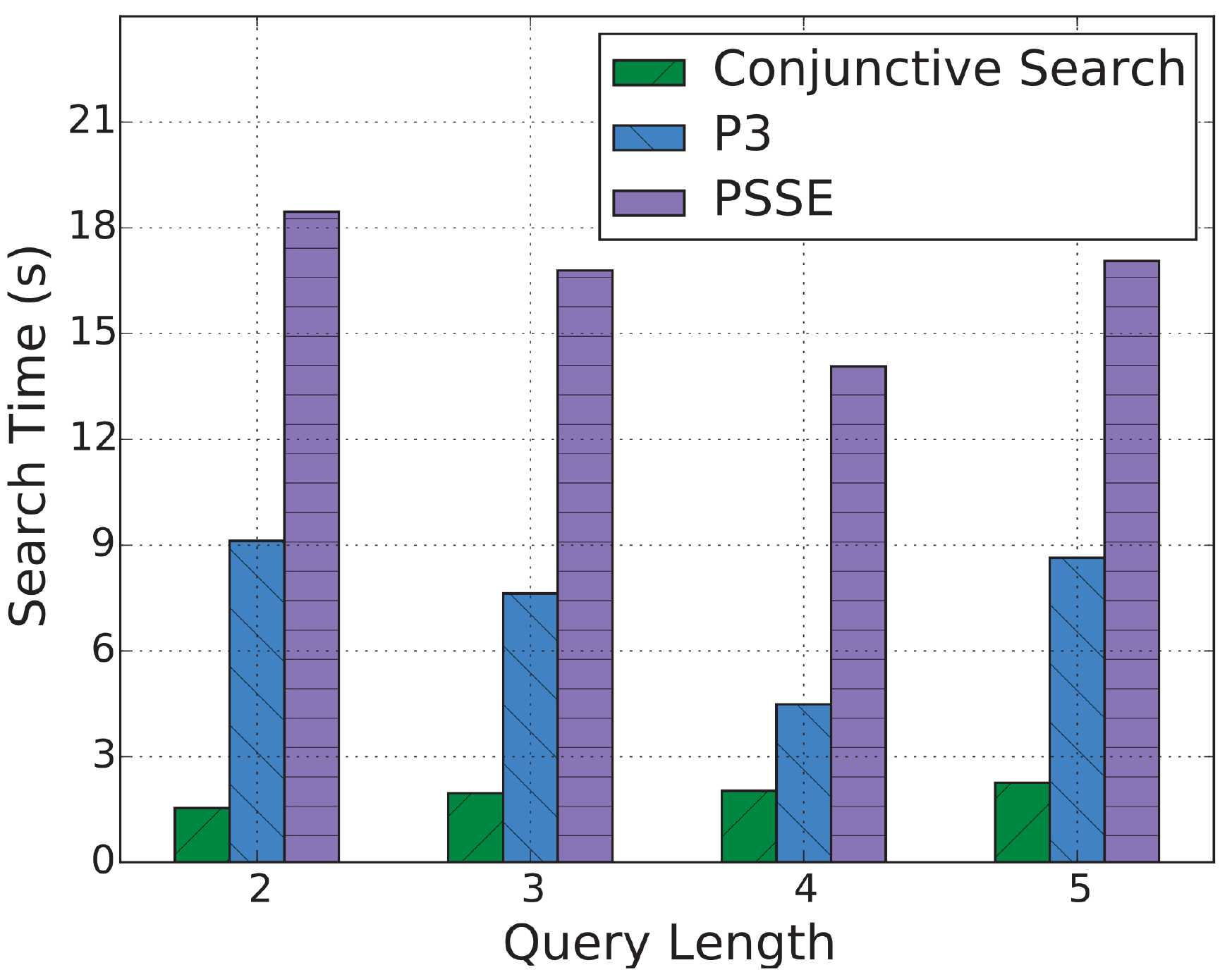}
    \caption{Search time for different query lengths ($m=2500$).}
    \label{fig:subfig_searchtime:a}
\end{minipage}%
\quad
\begin{minipage}[t]{0.32\textwidth}
    \centering
    \includegraphics[height=4.5cm]{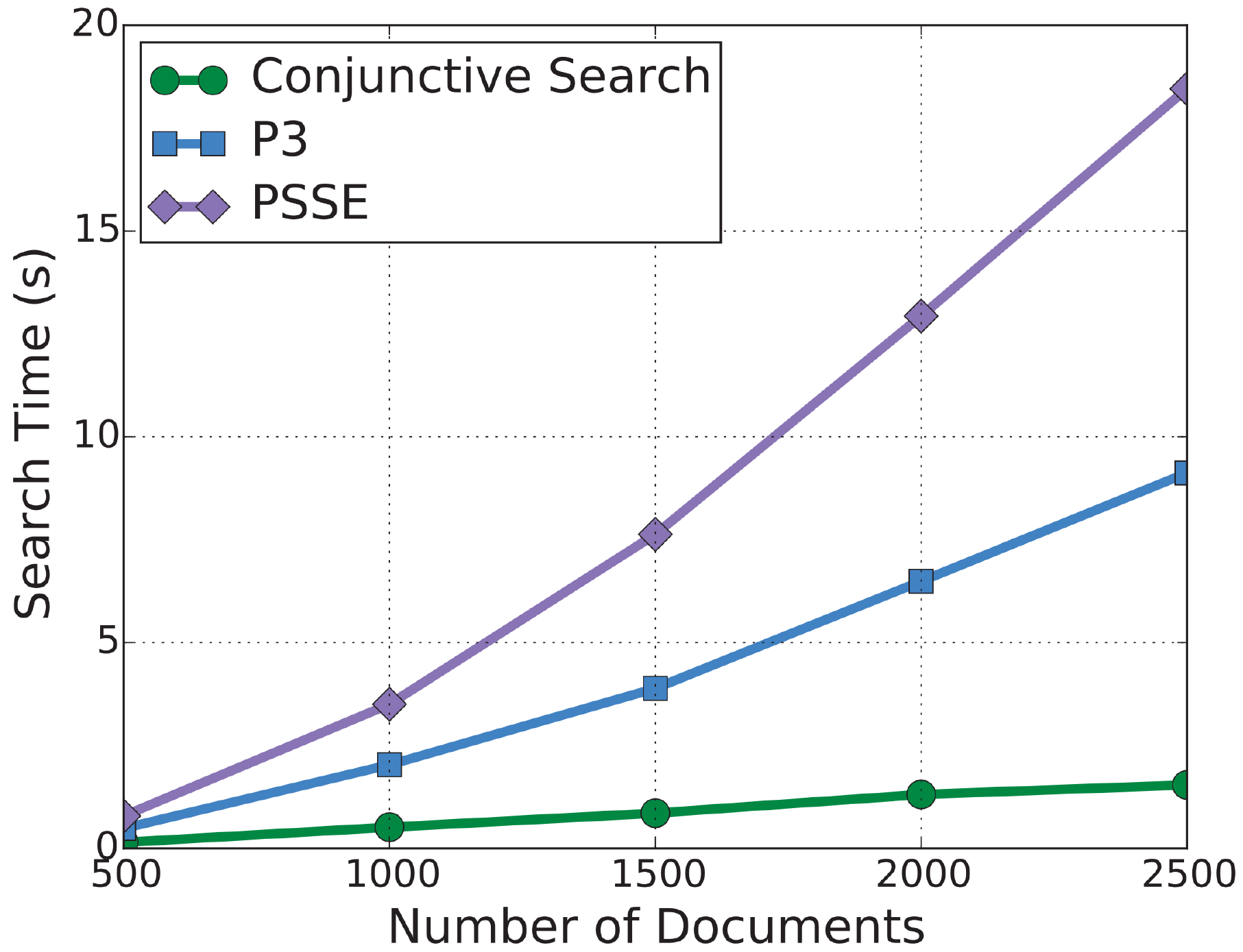}
    \caption{Search time for different numbers of indexed documents ($|Q|=2$).}
    \label{fig:subfig_searchtime:b}
\end{minipage}
\end{figure*}

%

\textbf{Query Time.}
The query time is defined as the time interval
from the submission of a user's trapdoor to
the receival of the search results.
For each queried phrase, we repeat the query $20$ times and calculate the average search time to mitigate the deviation caused by uncertain factors.
Note that PSSE may result in a huge index size (cf. Table \ref{tab:index}), which cannot be loaded completely into the memory used in our experiments.
Therefore, we enable the query algorithms of \texttt{P3} and PSSE to dynamically load the partial index.

Fig. \ref{fig:subfig_searchtime:a} shows the relationship between the search time and the query length.
The conjunctive search scheme takes the shortest search time.
However, such a scheme cannot provide accuracy guarantees as discussed in Section \ref{sec:search_accuracy}.
As to the phrase search schemes,
\texttt{P3} can roughly reduce the average search time of PSSE by half.
This is because PSSE has a large index size and thereby spends more time than \texttt{P3} on loading its index into the memory.


The search time with different document scales is shown in Fig. \ref{fig:subfig_searchtime:b}.
Here, we exhibit only the results for $|Q|=2$ due to space limitation.
The search time for each of the three methods enlarges with the growth of the number of documents.
Compared with PSSE, \texttt{P3} can greatly reduce the average search time for different scales of document sets.

\textbf{Communication overhead.}
The communication time and data volumes are depicted in Fig. \ref{fig:efficiency_communication}.
The communication time means the transmission time of the trapdoors and search results
between the client and the cloud server.
As the the number of indexed documents grows,
the communication time becomes higher for all three methods.
In particular, \texttt{P3} has the shortest communication time,
because $\texttt{P3}$ has the smallest data volume.
First, $\texttt{P3}$ has a higher search accuracy than the conjunctive search scheme, and thereby gets a smaller volume of search results that should be replied from the cloud server to the client.
Second, compared with PSSE, $\texttt{P3}$ only needs one-round of interaction and avoids sending intermediate data to the client for phrase recognition.


\begin{figure*}[t]
\begin{minipage}{0.5\textwidth}
  \centering
  \includegraphics[height=4.5cm]{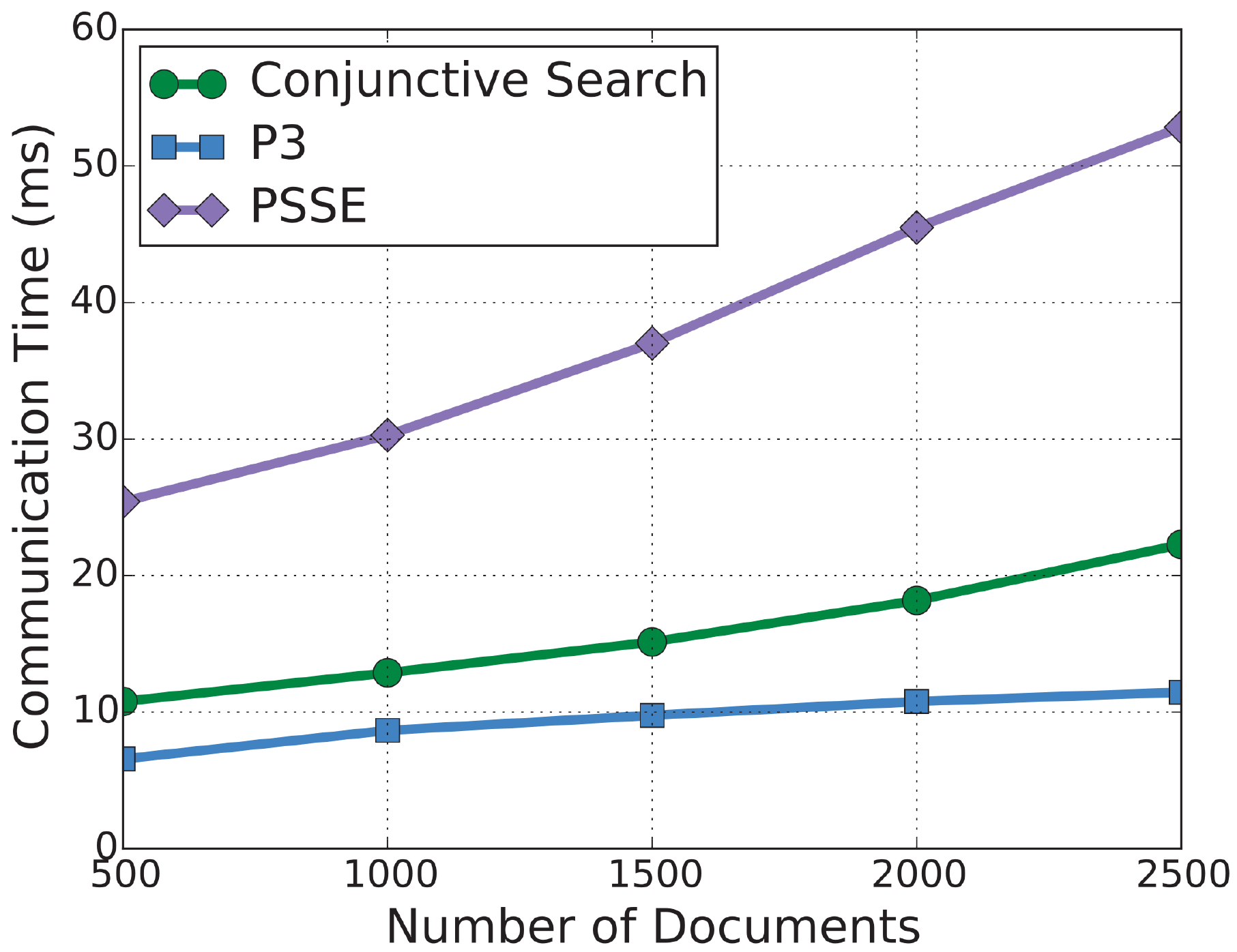}\\
  \scriptsize{(a) Communication time}
\end{minipage}
\quad
\begin{minipage}{0.5\textwidth}
  \centering
  \includegraphics[height=4.5cm]{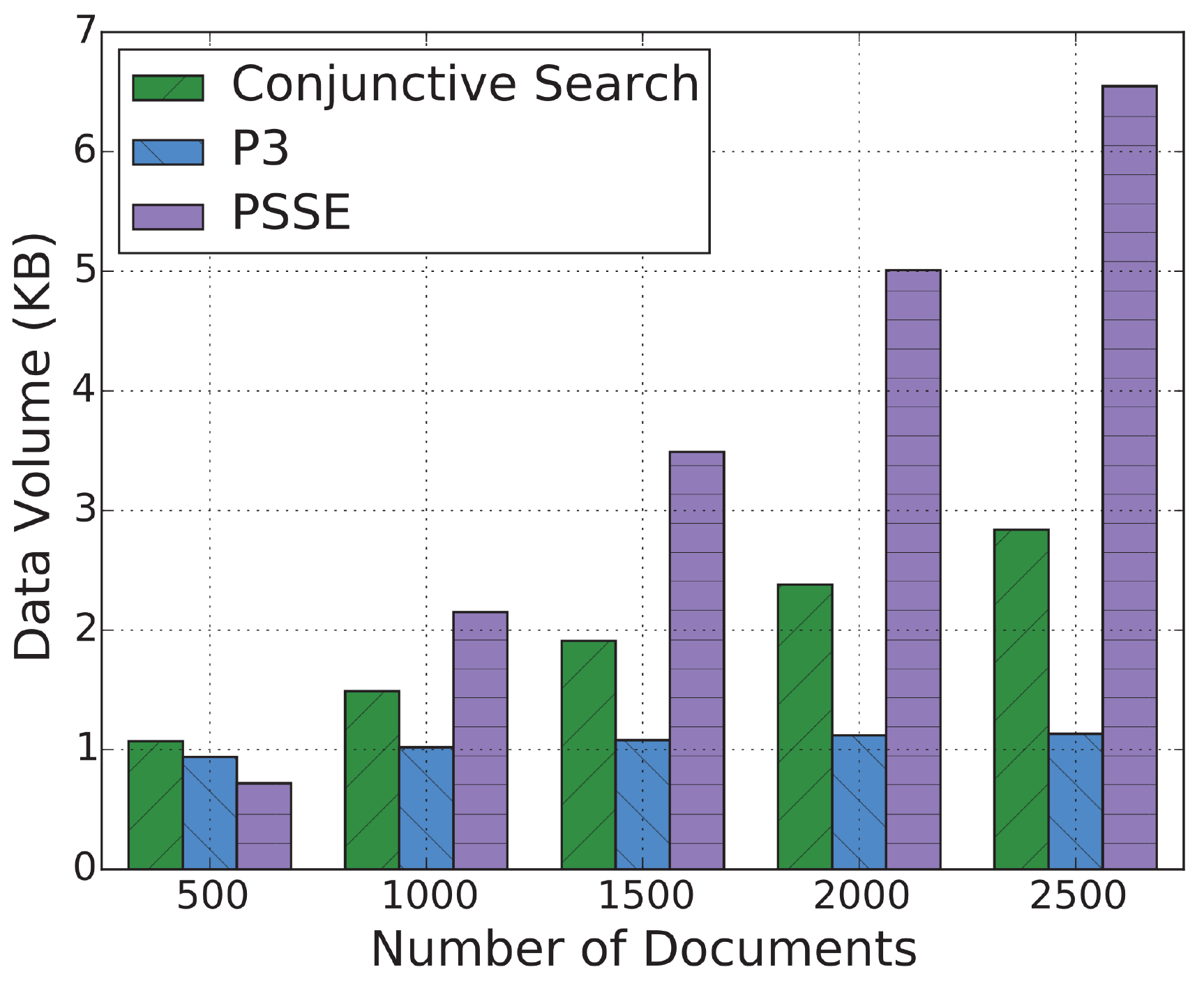}\\
  \scriptsize{(b) Communication data volumes}
\end{minipage}
\caption{Communication overhead with different numbers of documents ($|Q|=2$).}\label{fig:efficiency_communication}
\end{figure*}

\section{Discussion} \label{sec:discussion}
Although the proposed scheme is more efficient than the existing phrase search schemes,
there are still two limitations.

First, compared with the conjunctive search scheme,
\texttt{P3} has to spend more time on phrase recognition,
and thereby increases search time.
Second, \texttt{P3} cannot directly support a flexible index update due to the inherent feature of the inverted index and the adoption of the padding strategy.

A possible way to mitigate these limitations is leveraging the parallel processing techniques over server clusters.
We can partition the whole document set into several subsets, each of which contain partial documents and is indexed independently.
Given a phrase query from the user, search operations can be performed in parallel over the subsets, which helps to shorten the search time.
An offline update of the secure index can be employed to deal with updates, e.g., add or remove of documents and keywords.
In particular, when a document has to be updated, we only need to re-generate the index of the corresponding subset which the document belongs to, thereby reducing the index update overhead.
We leave these attempts for the future work.

\section{Conclusion}\label{sec:conclusion}
In this paper, we presented a novel scheme,
\texttt{P3}, which tackled the challenges
in phrase search for intelligent encrypted data processing in  cloud-based IoT.
The scheme exploits the homomorphic encryption and bilinear map to determine the pairwise location relationship of queried keywords on the cloud server side.
It eliminates the need of a trusted third party and greatly reduces communication overheads.
Thorough security analysis illustrated that the proposed scheme provides the desired security guarantees.
The experimental evaluation results demonstrated the effectiveness and efficiency of the proposed scheme.
In future work, we plan to further improve the flexibility and efficiency of the scheme.

\appendices




\ifCLASSOPTIONcaptionsoff
  \newpage
\fi

\bibliographystyle{abbrv}
\bibliography{main}

%
\vspace{-20pt}
\begin{IEEEbiographynophoto}
{Meng Shen} received the B.Eng degree from Shandong University, Jinan, China in 2009, and the Ph.D degree from Tsinghua University, Beijing, China in 2014, both in computer science. Currently he serves in Beijing Institute of Technology, Beijing, China, as an assistant professor. His research interests include privacy protection for cloud and IoT, blockchain applications, and encrypted traffic classification.
He received the Best Paper Runner-Up Award at IEEE IPCCC 2014.
He is a member of the IEEE.
\end{IEEEbiographynophoto}

\vspace{-30pt}

\begin{IEEEbiographynophoto}
{Baoli Ma} received the B.Eng degree in computer science from Beijing Institute of Technology, Beijing, China in 2015.
Currently he is a M.S. student in the Department of Computer Science, Beijing Institute of Technology.
His research interests include Cloud Computing and Secure Searchable Encryption.
\end{IEEEbiographynophoto}

\vspace{-30pt}
\begin{IEEEbiographynophoto}
{Liehuang Zhu} is a professor in the Department of Computer Science at Beijing Institute of Technology.
He is selected into the Program for New Century Excellent Talents in University from Ministry of Education, P.R. China.
His research interests include Internet of Things, Cloud Computing Security, Internet and Mobile Security.
\end{IEEEbiographynophoto}

\vspace{-30pt}

\begin{IEEEbiographynophoto}
{Xiaojiang Du} is a tenured professor in the Department of Computer and Information Sciences at Temple University, Philadelphia, USA. Dr. Du received his B.S. and M.S. degree in electrical engineering from Tsinghua University, Beijing, China in 1996 and 1998, respectively. He received his M.S. and Ph.D. degree in electrical engineering from the University of Maryland College Park in 2002 and 2003, respectively. His research interests are wireless communications, wireless networks, security, and systems. He has authored over 200 journal and conference papers in these areas, as well as a book published by Springer. Dr. Du has been awarded more than \$5 million US dollars research grants from the US National Science Foundation (NSF), Army Research Office, Air Force, NASA, the State of Pennsylvania, and Amazon. He won the best paper award at IEEE GLOBECOM 2014 and the best poster runner-up award at the ACM MobiHoc 2014.
He serves on the editorial boards of three international journals.
Dr. Du is a Senior Member of IEEE and a Life Member of ACM.
\end{IEEEbiographynophoto}

\begin{IEEEbiographynophoto}
{Ke Xu} received his Ph.D. from the
Department of Computer Science and Technology of
Tsinghua University, Beijing, China, where he serves
as a full professor. He has published more than 100
technical articles and holds 20 patents in the research areas of next generation Internet, P2P systems,
Internet of Things (IoT), network virtualization and
optimization. He is a member of ACM and has guest
edited several special issues in IEEE and Springer
Journals. Currently, he is holding visiting professor
position at University of Essex, UK.
\end{IEEEbiographynophoto}

\end{document}